\documentclass[12pt]{article}

\font\grande=cmr9.5 scaled \magstep4
\font\medio=cmr9.5 scaled \magstep2
\outer\def\beginsection#1\par{\medbreak\bigskip
      \message{#1}\leftline{\bf#1}\nobreak\medskip
\vskip-\parskip
      \noindent}
\usepackage{graphicx} 
\begin{document}
\bibliographystyle {unsrt}

\titlepage

\begin{flushright}
CERN-PH-TH/2013-051
\end{flushright}

\vspace{10mm}
\begin{center}
{\grande Bootstrapping from inflationary magnetogenesis}\\
\vspace{0.5cm}
{\grande to CMB initial conditions}\\
\vspace{1.5cm}
 Massimo Giovannini
 \footnote{Electronic address: massimo.giovannini@cern.ch}\\
\vspace{1cm}
{{\sl Department of Physics, 
Theory Division, CERN, 1211 Geneva 23, Switzerland }}\\
\vspace{0.5cm}
{{\sl INFN, Section of Milan-Bicocca, 20126 Milan, Italy}}
\vspace*{0.5cm}
\end{center}

\vskip 0.5cm
\centerline{\medio  Abstract}
\vskip 0.2cm
The temperature and polarization anisotropies of the Cosmic Microwave Background are analyzed under the hypothesis that the same inflationary seed accounting for protogalactic magnetism also affects the Einstein-Boltzmann hierarchy  whose initial conditions are assigned for typical correlation scales larger than the Hubble radius after matter-radiation equality but before decoupling. Since the primordial gauge spectrum depends on a combination of pivotal parameters of the concordance model, the angular power spectra of the temperature and of the polarization are computed, for the first time, in the presence of a putative large-scale magnetic field of inflationary origin and without supplementary hypotheses.
\vskip 0.5cm

\noindent

\vspace{5mm}

\vfill
\newpage
\renewcommand{\theequation}{1.\arabic{equation}}
\setcounter{equation}{0}
\section{Bootstrapping large-scale magnetism}
\label{sec1}
More than sixty years ago, an intense debate on the origin of galactic cosmic rays \cite{fermi,alfven} and 
the first ambiguous evidences of starlight polarization \cite{hiltnerhall,davis} led Fermi to propose the existence of a galactic magnetic field with approximate strength in the $\mu$G range\footnote{We shall employ hereunder the conventional prefixes of International System of units i.e. $\mu\mathrm{G} = 10^{-6}\, \mathrm{Gauss}$, $\mathrm{nG} = 10^{-9} \, \mathrm{Gauss}$ 
and so on and so forth.}. In the same period it was correctly argued that cosmic rays above $10^{13}$ eV were galactic (as opposed to solar) 
and the high degree of isotropy in the arrival directions could be explained by the presence of an irregular magnetic field 
of $\mu$G strength able to scramble the trajectories of charged species (see, for instance, \cite{cocconi} for a terse
account of this problem). 

Today we know, with a plausible degree of confidence, that  the galactic field  contains a large-scale regular component 
and a small-scale turbulent one, both having a local strength of a few $\mu$G (see, for instance, \cite{beck}). While the turbulent component dominates in strength by a factor of a few, the regular component (with approximate correlation scale between  $10$ to $30$ kpc) imprints dominant drift motions as soon as the Larmor radius of cosmic rays is larger than the maximal scale of the (kinetic or magnetic) turbulences which is ${\mathcal O}(100\,\mathrm{pc})$.  Clusters (i.e. gravitationally bound systems 
of galaxies)  have been shown to possess a large-scale magnetic field, in the $\mu$G range (more specifically between $10$ and $100$ nG) 
and with correlation scale larger than in the galactic case suggesting a quasi-flat magnetic power spectrum \cite{cl}. 
Superclusters (i.e. loosely bound systems of clusters) have been also claimed to have magnetic fields \cite{supclust} at the $\mu$G level even if, in this case, unresolved ambiguities persist on the way the magnetic field strengths are inferred from the Faraday rotation measurements.

In a seemingly different perspective we are now facing a problem conceptually close to the one of the early fifties of the past 
century. 
The spectrum of the highest energy cosmic rays shows no significant deviation from isotropy \cite{auger1}
as it can be argued from the distribution of arrival directions of cosmic rays detected above $10^{18}$ eV at the Pierre Auger Observatory. 
Purported correlations between the arrival directions of cosmic rays with energy above $6\times10^{19}$ eV and the positions of active galactic 
nuclei within $75$ Mpc \cite{auger2} are then statistically insignificant \cite{auger1}.  At smaller energies it has been 
convincingly demonstrated \cite{auger2} that overdensities on windows of $5$ deg radius (and for energies $10^{17.9} \mathrm{eV} < E < 10^{18.5} \mathrm{eV}$) are compatible with an isotropic distribution. We can then conclude that in the highest energy domain (i.e. energies larger  than about $60$ EeV),  cosmic rays are not appreciably deflected and basically isotropic.  As in the the early fifties the isotropy of the cosmic ray spectrum was a simple (thought indirect) evidence of galactic magnetism, it is tempting to speculate that 
the isotropy of the highest energy cosmic ray tells a similar story on irregular magnetic fields with correlation scales ${\mathcal O}(few\, \mathrm{Mpc})$ and within a cocoon of $70$ Mpc.

The current pieces of evidence confirm the early suggestions of Fermi arguing that large-scale magnetization should somehow be the remnant of the initial conditions of the gravitational collapse of the protogalaxy occurring over typical length-scales ${\mathcal O}(\mathrm{Mpc})$. Harrison \cite{har} and, with slightly different accents, Zeldovich \cite{zel} were among the first ones to propose that the source of magnetism over the largest length-scales could be searched in the pre-decoupling plasma and, since then, various mechanisms have been tailored for the generation of large-scale magnetic fields (see, for instance, \cite{review}). This class of problems has been dubbed some time ago magnetogenesis \cite{magn1}, i.e. the generation of large-scale magnetic fields during the early stages of the evolution of the plasma.  Since the magnetic fields must not jeopardize the spatial isotropy of the background geometry, their Fourier modes must be stochastically distributed and characterized by an appropriate power spectrum:
\begin{equation}
\langle {\mathcal B}_{i}(\vec{q},\tau)\, {\mathcal B}_{j}(\vec{k},\tau) \rangle = \frac{2\pi^2}{k^3}\, P_{{\mathcal B}}(k,\tau)\, P_{ij}(\hat{k})  \,\delta^{(3)}(\vec{q} + \vec{k}),
\label{av1}
\end{equation}
where, following the standard conventions (see e.g. \cite{magn2}), $P_{{\mathcal B}}(k,\tau)$ is the physical power spectrum, $P_{ij}(\hat{k})= ( \delta_{ij} - \hat{k}_{i}\,\hat{k}_{j} )$ and $\hat{k}_{i} = k_{i}/|\vec{k}|$. 
As it can be easily verified from the definition of the Fourier transform, $P_{{\mathcal B}}(k,\tau)$ 
has dimensions of an energy density and its square root has, therefore, the dimensions of a field intensity. 
Since magnetic fields exist over 
increasing length-scales with almost the same intensity, the parametrization of  Eq. (\ref{av1}) implies that 
$P_{{\mathcal B}}(k,\tau)$ can be considered, in the first approximation, as nearly scale-invariant. Incidentally, on a
technical ground, the parametrization of Eq. (\ref{av1}) follows the same conventions employed when the power spectra of curvature 
perturbations are assigned at the standard pivot scale $k_{\mathrm{p}}= 0.002\,\, \mathrm{Mpc}^{-1}$; denoting with 
${\mathcal R}(\vec{k},\tau)\simeq {\mathcal R}_{*}(\vec{k})$ the constant mode of curvature perturbations on comoving orthogonal hypersurfaces prior to equality, its two 
point function in Fourier space is 
\begin{equation}
\langle {\mathcal R}(\vec{q})\, {\mathcal R}(\vec{k}) \rangle = \frac{2\pi^2}{k^3}\, P_{{\mathcal R}}(k)\,\delta^{(3)}(\vec{q} + \vec{k}),\qquad P_{{\mathcal R}}(k) = {\mathcal A}_{{\mathcal R}} \biggl(\frac{k}{k_{\mathrm{p}}}\biggr)^{n_{\mathrm{s}}-1},
\label{av1a}
\end{equation}
where, using the WMAP 9yr data alone \cite{wmap1,wmap2,wmap3} in the light of the concordance scenario ${\mathcal A}_{{\mathcal R}}=
(2.41\pm 0.10)\times 10^{-9}$ and $n_{\mathrm{s}} = 0.972\pm0.013$. As in the magnetic case, the exact scale-invariant limit 
is realized when $n_{\mathrm{s}}\to 1$.

Magnetic fields can be produced during a phase of decelerated expansion inside the particle horizon which 
always grows much faster than the correlation scale of the field. It is then unlikely to obtain, in this case, correlation scales 
${\mathcal O}(\mathrm{Mpc})$ at the onset of protogalactic collapse: this is one of the main drawbacks of what is often called, somehow improperly,  
causal magnetogenesis \cite{magn2}. Conversely during inflation the Hubble radius is almost constant ${\mathcal O}(H^{-1})$ and it roughly
coincides with the event horizon. The correlation scale of a putative magnetic field amplified during inflation evolves much more rapidly than 
the Hubble radius itself so that the magnetic power spectrum at the end of inflation has an approximate amplitude ${\mathcal O}(H^4)$.
In the context of inflationary models the correlation length-scales of the produced magnetic fields range between ${\mathcal O}(k_{\mathrm{p}}^{-1})$ and ${\mathcal O}(k_{\mathrm{L}}^{-1})$ where $k_{\mathrm{p}}$ (see Eq. (\ref{av1a})) is the comoving scale at which the power spectrum of curvature perturbations is assigned in standard analyses and $k_{\mathrm{L}} \simeq {\mathcal O}(\, \mathrm{Mpc}^{-1})$ corresponds to the wavenumber of the gravitational collapse of the protogalaxy. 

To make any statement concerning the early evolution of the plasma, the post-inflationary dynamics must be specified within a reasonable degree 
of accuracy.  Direct Cosmic Microwave Background observations \cite{wmap1,wmap2,wmap3} (CMB in what follows), light curves of type-Ia supernovae \cite{SN1,SN2} and large-scale galaxy surveys \cite{SDSS} are compatible with a wide class of scenarios arranged around the 
$\Lambda$CDM paradigm where $\Lambda$ stands for the dark energy component 
and CDM for the cold dark matter component. The $\Lambda$CDM paradigm consists, nominally, of 
$6$ independent parameters which are extracted from the observational data. However, 
definite statements demand also more stringent assumptions 
on the origin of the  adiabatic mode or on the thermal history of the plasma at least 
up to  energy scales as large as few TeV. A conventional completion of the $\Lambda$CDM paradigm consists 
of a phase  of slow-roll inflation with standard thermal history where the post-inflationary evolution is suddenly 
dominated by radiation down to the scale of matter-radiation equality. In the latter framework the CMB scale 
left the inflationary Hubble radius around $65$ efolds prior to the end of inflation; conversely the scale of the 
protogalactic  collapse left the inflationary Hubble radius around $9$ efolds after the CMB scale.
 
The conventional completion of the $\Lambda$CDM paradigm implies that the typical wavenumber of the gravitational collapse of the protogalaxy 
in units of the inflationary Hubble rate is: 
\begin{equation}
\frac{k_{\mathrm{L}}}{a H} = 3.22\times 10^{-25} \, \biggl(\frac{k_{\mathrm{L}}}{\mathrm{Mpc}^{-1}}\biggr) \, \biggl(\frac{\epsilon}{0.01}\biggr)^{-1/4} \, 
\biggl(\frac{{\mathcal A}_{\mathcal R}}{2.41\times 10^{-9}} \biggr)^{-1/4},
\label{ex}
\end{equation}
where, besides ${\mathcal A}_{\mathcal R}$,  $\epsilon= - \dot{H}/H^2$ defines the standard slow-roll
parameter. The quasi flat spectrum of inflationary origin at the present time and for a typical comoving scale 
${\mathcal O}(k_{\mathrm{L}})$ can then be fully specified in terms of the standard thermal history and it is given by\footnote{Note that $h_{0}^2 \Omega_{\mathrm{R}}$ corresponds to the 
present critical fraction of radiation given as the sum of the energy density of the photons and of the neutrinos which are 
strictly massless in the $\Lambda$CDM scenario. }
\begin{equation}
P_{{\mathcal B}}(k_{\mathrm{L}},\tau_0) = 10^{-2.43} \, \biggl(\frac{{\mathcal A}_{{\mathcal R}}}{2.41 \times 10^{-9}}\biggr)\,
\biggl(\frac{\epsilon}{0.01}\biggr)\, \biggl(\frac{h_{0}^2 \Omega_{\mathrm{R}0}}{4.15 \times 10^{-5}}\biggr)\,  \biggl(\frac{r}{0.22}\biggr)\,\, \mathrm{nG}^2, 
\label{sp1}
\end{equation}
where it has been assumed  $P_{{\mathcal B}}(k,\tau_{\mathrm{end}}) = r H^4 \simeq {\mathcal O}(H^4)$ at the end of inflation\footnote{In specific models it turns out that the power spectrum, in the quasi-flat case, is given by $r\, H^4$ where $r$ is a numerical factor ranging between $0.1$ and $0.01$. The reference value of $r$ is $r= 9/(4\pi^2) \simeq 0.22$ and it corresponds to the one obtainable in the context of explicit models mentioned later on.}. The magnetic field itself can by estimated by taking the square root of the Eq. (\ref{sp1}) giving a field of the order of $0.06$ nG.  Equation (\ref{sp1}) is compatible with a galactic magnetic field of the order of the $\mu$G since 
during the process of collapse (prior to the onset of galactic rotation) the magnetic flux is frozen into the plasma element thanks to the large value of the conductivity. The mean matter density increases, during collapse,
from its critical value (i.e. $\rho_{\mathrm{cr}} =1.05\times 10^{-5} \,h_{0}^2 \mathrm{GeV}/\mathrm{cm}^3$) to a final value $\rho_{\mathrm{f}}$ value which is 5 to 6 orders of magnitude larger than $\rho_{\mathrm{c}}$. The magnetic power spectrum after collapse will then be  larger than (\ref{sp1}) by a factor  which is roughly $(\rho_{\mathrm{f}}/\rho_{\mathrm{c}})^{4/3}$. 

 As a consequence of the standard thermal history, the amplitude of the magnetic power spectrum depends only on
 $\epsilon$ and ${\mathcal A}_{{\mathcal R}}$ and the inflationary Hubble rate in Planck units 
is given by $H = \sqrt{\pi \, \epsilon\, {\mathcal A}_{{\mathcal R}}} \, M_{\mathrm{P}}$.  But since ${\mathcal A}_{{\mathcal R}}$ and $h_{0}^2 \Omega_{\mathrm{R}}$ are fixed within the vanilla $\Lambda$CDM scenario, the amplitude of the seed spectrum effectively depends only on $\epsilon$.
The purpose of this paper is to determine the initial conditions of the Einstein-Boltzmann hierarchy in terms of the early initial conditions derivable 
in the framework of inflationary magnetogenesis. It has been pointed out (see \cite{magn3} and references therein) that a key role in the analysis of large-scale magnetism must rely on an accurate determination of the distortions induced on the temperature and polarization anisotropies\footnote{In the $\Lambda$CDM paradigm the temperature and polarization anisotropies are given in terms of the temperature autocorrelations (the TT power spectrum), the polarization autocorrelations (the EE power spectrum) and the temperature-polarization cross-correlations (the TE power spectrum). The other 
correlations are vanishing even if, in the presence of a stochastic magnetic field, a B-mode power spectrum is naturally induced by Faraday rotation.} and various results have  been derived both analytically and numerically but never in terms of a primordial inflationary seed. 
The aim here is to bridge this gap by treating large-scale fields not generically present prior to decoupling but rather coming from inflationary magnetogenesis and to derive the initial conditions of the Einstein-Boltzmann hierarchy.  The TT, EE and TE angular power spectra will then be directly computed in terms of the appropriate initial conditions.

To develop a theory of the magnetized initial conditions without running into well known troubles, an efficient tool is the synchronous gauge which is particularly suitable for the perturbative description of the anisotropic stresses. The synchronous results will be explicitly connected to the set of gauge-invariant variables derived by Lukash \cite{lukash} in the context of the Lifshitz formalism \cite{lif} (see also \cite{strokov,luk2}). During the post-inflationary epoch, the evolution of the whole irrotational system is reduced 
to a single  normal mode which is invariant under infinitesimal coordinate transformations (as required in the context of the  Bardeen formalism \cite{bard1}) and whose source terms depend on the magnetic inhomogeneities. In the absence of magnetic fields, earlier analyses  \cite{KS,chibisov} followed the same logic of \cite{lukash} but in the case of scalar field matter.  We shall use that the normal modes identified in \cite{lukash,KS,chibisov}  are all related to the (rescaled) curvature perturbations on comoving orthogonal hypersurfaces \cite{br1,bard2}.

The layout of the present analysis is as follows.
In section \ref{sec2} the fluctuations of inflationary magnetogenesis are analyzed in the synchronous 
coordinate system. In section \ref{sec3} the initial conditions of the Einstein-Boltzmann hierarchy are bootstrapped out of the values provided by inflationary magnetogensis by using the explicit evolution of curvature perturbations for typical scales larger than the Hubble radius at the corresponding epoch. The whole 
system of perturbations is then integrated across the radiation-matter transition. In section \ref{sec4}  the TT, EE and TE correlations are explicitly computed in the case of a magnetized adiabatic mode of inflationary origin. The B-mode autocorrelations arising in this approach are also 
briefly examined. Section \ref{sec5} contains the concluding remarks. In the appendix some of technical results have been collected for the interested reader to avoid excessive digressions.
\renewcommand{\theequation}{2.\arabic{equation}}
\setcounter{equation}{0}
\section{Inflationary evolution}
\label{sec2}
During a phase of slow-roll dynamics\footnote{Consistently with the absence of spatial curvature of the 
$\Lambda$CDM paradigm, the  background geometry is assumed to be conformally flat throughout this investigation; the line element is given by $ds^2 = a^2(\tau)[d\tau^2 - d\vec{x}^2]$. The prime denotes
a derivation with respect to the conformal time coordinate $\tau$ while the overdot indicates a derivation with respect to the cosmic time coordinate $t$. The conformal time coordinate $\tau$ is related to the cosmic time as $a(\tau) d\tau = dt $.} magnetic fields are amplified while the electric fields are suppressed 
thanks to the coupling to a spectator field \cite{spec1,spec2,mgcond}. The electric and magnetic power spectra contribute to the  fluctuations of the various components of the energy-momentum tensor and, ultimately, 
to the curvature perturbations in the same way as a putative magnetic field affects curvature 
perturbations across the matter radiation transition (see \cite{magn2} and references therein). The form of the electric and magnetic power spectra will now be discussed not with the aim of endorsing a 
particular magnetogenesis scenario but with the purpose of drawing general lessons on the parametrization of the magnetized power spectra of curvature perturbations at the end of inflation.
\subsection{Parametrization of the power spectra}
Consider an explicit coupling of the Abelian kinetic term to a spectator field $\sigma$:
\begin{eqnarray}
S &=& \int d^{4} x\, \sqrt{ -g} \, \, \biggl[ - \frac{1}{2 \ell_{\mathrm{P}}^2} R + \frac{1}{2}  
g^{\alpha\beta} \partial_{\alpha} \varphi \partial_{\beta} \varphi - V(\varphi) 
\nonumber\\
&+& \frac{1}{2}  
g^{\alpha\beta} \partial_{\alpha} \sigma \partial_{\beta} \sigma - W(\sigma)  - \frac{\lambda(\sigma)}{16\pi} \, Y_{\mu\nu} \, Y^{\mu\nu}  - \,j_{\mu} \, Y^{\mu} \biggr],
\label{action1}
\end{eqnarray}
where $\varphi$ denotes the inflaton field, $Y_{\mu}$ the gauge field, $Y_{\mu\nu}$ the gauge field 
strength; $j_{\mu}$ represents the potential contribution of an Ohmic current. 
The evolution of the gauge fields in this class of models 
has been analyzed in different situations (see, e.g. \cite{spec1,spec2} and references therein). If $\lambda(\sigma)$ is a monotonically increasing function of the conformal time coordinate $\tau$, magnetic fields are amplified while the electric fields are suppressed. This kind of monotonic behavior is quite natural in conventional inflationary models where the curvature scale increases as we approach the protoinflationary stage which is customarily modeled with a phase of decelerated expansion preceding the ordinary slow-roll phase.  During the protoinflationary phase conformal invariance can still be unbroken and, if this happens, initial Ohmic currents (remnants of the protoinflationary dynamics) are not dissipated \cite{mgcond}. Conducting initial conditions may then dominate against the quantum initial conditions and the increase of $\lambda$ corresponds, in this case, to an increase of the Debye shielding length and 
to a further suppression of the electric fields \cite{mgcond}.  For an  exponential coupling we shall have, for instance, $\ln{\lambda(\sigma)} = \gamma_{\sigma} \sigma/M$ where $M$ is a typical scale characterizing the evolution of the spectator field.  
The equations of motion of $\sigma$ can be solved at the level of the background and more specific situations can be found, for instance, in \cite{spec1}
(see also \cite{spec2}).  

In a conformally flat background geometry characterized by a scale factor $a(\tau)$ in the conformal time parametrization, the power spectra of the comoving electric and magnetic fields are\footnote{Note that $H_{\nu}(z)$ are standard Hankel functions. Furthermore $|{\mathcal N}|^2= \pi/2$.}:
\begin{eqnarray}
&& P_{\mathrm{B}}(k,\tau) = \frac{k^5}{2\pi^2} |f_{k}(\tau)|^2,\quad f_{k}(\tau) =  \frac{{\mathcal N}}{\sqrt{2 k}} \, \sqrt{- k \tau} \, H_{\nu}^{(1)}(- k \tau), 
\label{PB}\\
&& P_{\mathrm{E}}(k,\tau) = \frac{k^3}{2\pi^2} |g_{k}(\tau)|^2,\quad g_{k}(\tau) = - {\mathcal N}\, \sqrt{\frac{k}{2}}\, \sqrt{-k\tau} \, H_{\nu-1}^{(1)}(- k \tau),
\label{PE}
\end{eqnarray}
where the mode functions $f_{k}(\tau)$ and $g_{k}(\tau)$ are solutions of the corresponding equations\footnote{The presence of the conductivity 
$\sigma_{\mathrm{c}}$ (not to be confused with the spectator field) plays a specific role during reheating and in the case of conducting initial conditions. More details on these issues can be found in \cite{mgcond}. Here we shall focus, for illustration, on the conventional case of quantum mechanical initial conditions, as it can be 
argued from the boundary conditions imposed on the mode functions.}:
\begin{equation}
f_{k}' = g_{k} + {\mathcal F} f_{k},\qquad 
g_{k}' = - k^2 f_{k} - {\mathcal F} g_{k} - 4 \pi \sigma_{\mathrm{c}} g_{k}.
\label{fg}
\end{equation}
In Eq. (\ref{fg}) the prime denote a derivation with respect to the conformal time coordinate and ${\mathcal F} = \sqrt{\lambda}'/\sqrt{\lambda}$. To keep the discussion general without selecting a specific model,
the growth rate ${\mathcal F}$ is parametrized as ${\mathcal F} = f {\mathcal H}$ with
${\mathcal H}= a H$; in the latter case the Bessel index appearing in Eqs. (\ref{PE}) and (\ref{PB}) is $\nu = 1/2 + f + f \epsilon$ and $\epsilon$ is the slow-roll parameter entering the expression of $\nu$ since the relation between the Hubble rate and the conformal time coordinate demands the following well known condition $(1 - \epsilon) aH= -\tau$. In the case of Eq. (\ref{action1}) 
$f = \epsilon\, \gamma_{\sigma}/(1-\epsilon)$ but the parametrization in terms of the growth rate has the advantage of being sufficiently general to incorporate at once different dynamical situations. The quasi-flat magnetic spectrum corresponds to the case $f \simeq 2$; direct analyses of these scenarios show that, in this class of models, the departure from scale-invariance cannot be too large (i.e. $f \leq 2.2$) \cite{mgfluc} if the adiabatic mode is to be the dominant source of inhomogeneities across matter-radiation decoupling. After the end of inflation ${\mathcal F}\to 0$ and the relation between the physical and the comoving power spectra is given by $P_{\mathrm{B}}(k,\tau) = a^4(\tau) P_{\mathcal{B}}(k,\tau)$ and $P_{\mathrm{E}}(k,\tau) = a^4(\tau) P_{\mathcal{E}}(k,\tau)$. 

As discussed in appendix \ref{APPA} the fields $\vec{B}(\vec{x}, \tau)$ and $\vec{E}(\vec{x},\tau)$ correspond to the canonical 
normal modes of the system diagonalizing the Hamiltonian density and having well defined 
properties of transformation under the duality symmetry \cite{mgcond,duality1,duality2}.  In terms of the power spectra of the comoving magnetic and electric fields the correlation functions are given by 
\begin{eqnarray}
&& \langle B_{i}(\vec{q},\tau)\, B_{j}(\vec{p},\tau) \rangle = \frac{2\pi^2}{q^3}\, P_{\mathrm{B}}(q,\tau)\, P_{ij}(\hat{q})  \,\delta^{(3)}(\vec{q} + \vec{p}),
\label{st1}\\
&& \langle E_{i}(\vec{q},\tau)\, E_{j}(\vec{p},\tau) \rangle = \frac{2\pi^2}{q^3}\, P_{\mathrm{E}}(q,\tau)\, P_{ij}(\hat{q})  \,\delta^{(3)}(\vec{q} + \vec{p}).
\label{st2}
\end{eqnarray}
In full analogy with the power spectra of curvature perturbations introduced in section \ref{sec1}, the magnetic power spectra can be expressed as 
\begin{eqnarray}
&&P_{\mathcal B}(k,\tau) = A_{B}(k_{\mathrm{L}}, \tau) \biggl(\frac{k}{k_{\mathrm{L}}}\biggr)^{n_{\mathrm{B}}-1},\qquad n_{\mathrm{B}} = 5 - 2 f - 2 f \epsilon,
\nonumber\\
&& A_{B}(k_{\mathrm{L}},\tau) = \frac{9\, H^4}{4\pi^2} \,\biggl(\frac{a_{1}}{a}\biggr)^{4} \,\biggl(\frac{k_{\mathrm{L}}}{a H} \biggr)^{n_{\mathrm{B}} -1} \, {\mathcal K}(n_{\mathrm{B}}),\qquad {\mathcal K}(n_{\mathrm{B}})= \frac{2^{5 - n_{\mathrm{B}}}}{9 \pi}\,\Gamma^2\biggl(\frac{6- n_{\mathrm{B}}}{2}\biggr),
\label{sp1a}
\end{eqnarray}
where $(k_{\mathrm{L}}/aH)$, as already stressed after Eq. (\ref{ex}), in the $\Lambda$CDM scenario, is solely determined in terms of $\epsilon$ and ${\mathcal A}_{{\mathcal R}}$. 
In Eq. (\ref{sp1a}) ${\mathcal K}(n_{\mathrm{B}})$ varies very little\footnote{Note that  ${\mathcal K}(1) =1$.} in the range of physical rates $2\leq f < 2.2$; for values $f > 2.2$ (excluded from the present considerations) the energy density fluctuations induced by the magnetic fields will get larger than the adiabatic mode. The amplitude 
$A_{B}(k_{\mathrm{L}},\tau)$ depends on the thermal history through the factor 
$(a_{1}/a)$ which gets different values depending on the epoch at which the power spectrum is evaluated, for instance, at equality,
\begin{equation}
\biggl(\frac{a_{1}}{a_{\mathrm{eq}}}\biggr)  = \biggl(\frac{2 \Omega_{\mathrm{R}0}}{\pi \epsilon {\mathcal A}_{{\mathcal R}} }\biggr)^{1/4} \, \sqrt{\frac{H_{0}}{M_{\mathrm{P}}}} \biggl(\frac{\Omega_{\mathrm{M}0}}{\Omega_{\mathrm{R}0}}\biggr)
\label{sp2a}
\end{equation}
where $h_{0}^2 \Omega_{\mathrm{R}0} = 4.15\times 10^{-5}$ and the total matter fraction at the 
present time is $\Omega_{\mathrm{M}0} = \Omega_{\mathrm{c}0} + \Omega_{\mathrm{b}0}$ is given 
by the sum of the CDM and of the baryonic contribution. Following the same logic of Eq. (\ref{sp2a}), but at a different redshift, Eq. (\ref{sp1}) can be derived from Eq. (\ref{sp1a}) by recalling known conversion 
factors between different system of units\footnote{It is useful, for numerical estimates, to appreciate that $M_{\mathrm{P}}^2 = 2.149\times 10^{66} \,\,\mathrm{nG}$.}.

The parameters of  Eq. (\ref{sp2a}) can all be extracted, assuming the $\Lambda$CDM model, by using different data sets \cite{wmap1,wmap2,wmap3} (see also e.g. Eqs. (\ref{ppp1})--(\ref{ppp3}) in section \ref{sec4}). The only 
exception is represented by  $\epsilon$ which cannot be determined in the vanilla $\Lambda$CDM model but it can be bounded from above in terms the tensor to scalar ratio $r_{\mathrm{T}}= {\mathcal P}_{\mathrm{T}}/{\mathcal P}_{{\mathcal R}}$ measuring the ratio between the spectrum of the tensor modes and the spectrum of the curvature perturbations at the pivot scale $k_{\mathrm{p}}$.  To first-order 
in the slow-roll expansion the tensor to scalar ratio $r_{\mathrm{T}} = 16 \epsilon + {\mathcal O}(\epsilon^2)$ so that any bound on $r_{\mathrm{T}}$ can be translated into a bound on $\epsilon$.  
\begin{table}[!ht]
\begin{center}
\begin{tabular}{||c|c|c|c|c|c||}
\hline
Data & $r_{\mathrm{T}}$ &$n_{\mathrm{s}}$ & $\epsilon$\\
\hline
 WMAP9 alone& $<0.38$ &$ 0.992\pm 0.019$&$<0.023$\\
WMAP9 + Hubble&  $<0.34$ &$ 0.995\pm 0.015$&$<0.021$\\
WMAP9+ BAO &$<0.18$ &$ 0.973\pm0.011$&$<0.011$\\
WMAP9+ all& $<0.13$ & $0.9647_{-0.0084}^{0.0083}$&$ <0.0081$\\
\hline
\end{tabular}
\caption{The upper limits on the tensor-to-scalar ratio and on the 
slow-roll parameter for some illustrative choices of cosmological data sets.}
\label{TABLE1}
\end{center}
\end{table}
The  WMAP9 observations can be combined with different data sets in the light 
of the $\Lambda$CDM scenario supplemented by the tensor modes 
of the geometry. In this way it is possible to obtain a bound on $r_{\mathrm{T}}$ and 
on $\epsilon$. In Tab. \ref{TABLE1} the bound on $r_{\mathrm{T}}$ obtainable 
in the case of the WMAP9 data alone is compared with the same bound obtained 
by combining the WMAP9 data with the measurements on the Hubble constant 
of Ref. \cite{perci} or by combining the WMAP9 observations with the baryon acoustic 
oscillations, dubbed BAO in Tab. \ref{TABLE1} (see e.g. \cite{SDSS}). The last row of Tab. \ref{TABLE1} refers to the combination of the WMAP9 data with almost all the data sets recently available, namely, the data on the Hubble rate, the ones on the baryon acoustic oscillations supplemented by the ones of the Atacama Cosmology Telescope \cite{act}, by the data of the south pole telescope \cite{spt} and by the three year sample of the supernova legacy survey \cite{snls3}.

For the standard thermal history with sudden reheating also the conductivity $\sigma_{\mathrm{c}}$ jumps at a finite value at the end of inflation and the continuity of the electric and magnetic fields
 implies that the amplitude of the electric power spectrum gets 
suppressed, at a fixed time, as $(k/\sigma_{\mathrm{c}})^2$ in comparison with its magnetic counterpart \cite{mgcond}. The electric mode functions are exponentially suppressed, for a fixed wavenumber. The magnetic power spectrum is also suppressed, for 
sufficiently large $k$, as $\exp{[-  2(k^2/k_{\sigma}^2)]}$ 
 where $k_{\sigma}^{-2}  = \int_{\tau_{\sigma}}^{\tau} \, d\tau' /[4\pi \sigma_{\mathrm{c}}(\tau')]$. The evaluation 
of $k_{\sigma}$ is complicated by the fact that the integral extends well after $\tau_{\sigma}$. This estimate
can be made rather accurate by computing the transport coefficients of the plasma in different regimes \cite{cond}.  By taking $\tau = \tau_{\mathrm{eq}}$
\begin{equation}
\biggl(\frac{k}{k_{\sigma}}\biggr)^2 = \frac{4.75 \times 10^{-26}}{ \sqrt{2 \, h_{0}^2 \Omega_{\mathrm{M}0} (z_{\mathrm{eq}}+1)}} \, \biggl(\frac{k}{\mathrm{Mpc}^{-1}} \biggr)^2,
\label{CC4}
\end{equation}
showing that $\exp{[ - 2(k/k_{\sigma})^2]}$ is so small to give negligible suppression for ${\mathcal O}(k_{\mathrm{p}}) \leq k \leq {\mathcal O}(k_{\mathrm{L}})$ where the present considerations apply.  

\subsection{Energy density and anisotropic stress}
The normalized fluctuation of the magnetic energy density and of the related anisotropic stress are given by:
\begin{equation}
\delta \rho_{\mathrm{B}}(\vec{x},\tau) =\int \frac{d^{3} q}{(2\pi)^{3/2}} \, \delta\rho_{\mathrm{B}}(\vec{q},\tau)\, e^{- i \vec{q}\cdot \vec{x}},\quad\Pi_{ij}^{(\mathrm{B})}(\vec{x},\tau) = \int\frac{d^{3} q}{(2\pi)^{3/2}} \Pi^{(B)}_{ij}(\vec{q},\tau)e^{- i \vec{q}\cdot\vec{x}},
\label{en1}
\end{equation}
The normalized fluctuation of the magnetic pressure is $\delta p_{\mathrm{B}}(\vec{x},\tau) = \delta\rho_{\mathrm{B}}(\vec{x},\tau)/3$.  Since we shall be dealing with the scalar modes of the geometry, it is practical to introduce the scalar projections of the anisotropic stress
\begin{equation}
\nabla^2 \Pi_{\mathrm{B}}(\vec{x},\tau) = \partial_{i} \partial_{j} \Pi^{ij}_{\mathrm{B}}(\vec{x},\tau).
\label{en6}
\end{equation}
The fluctuations of the energy density and of the anisotropic stress in Fourier space are reported in appendix \ref{APPA} to avoid lengthy digressions since this analysis  is standard and can be found within slightly different perspectives  in \cite{magn2,magn3,mgfluc}. In full analogy with the magnetic fluctuations the fluctuations of the 
electric energy density and of the electric anisotropic stress can be defined.  

The magnetic energy density and and the anisotropic stress induce scalar fluctuations 
of the geometry as established long ago (see \cite{magn2} and references therein).  
The variables $h(k,\tau)$ and $\xi(k,\tau)$ parametrize the metric fluctuations in the 
synchronous coordinate system 
\begin{equation}
\delta_{\mathrm{s}} g_{i\,j}(k,\tau) = a^2 \biggl[ \hat{k}_{i} \hat{k}_{j} h + 6 \xi\biggl( \hat{k}_{i} \hat{k}_{j} - \frac{\delta_{ij}}{3}\biggr)\biggr],\quad \delta_{\mathrm{s}} g_{0\,i}(k,\tau)= \delta_{\mathrm{s}} g_{00}(k,\tau)= 0,
\label{metric}
\end{equation}
where $\delta_{\mathrm{s}}$ stresses that we are here considering the scalar modes of the geometry. 
For practical reasons, in what follows, this subscript will be omitted. 

The evolution of $h(k,\tau)$ and $\xi(k,\tau)$ can be obtained by perturbing the Einstein equations during the inflationary phase:
\begin{eqnarray}
&&-  2 k^2 \xi + {\mathcal H} h' = - 4\pi G a^2 (\delta\rho_{\varphi} + \delta\rho_{\sigma} + \delta\rho_{\mathrm{B}}),
\label{one}\\
&& h'' + 2 {\mathcal H} h' - 2 k^2 \xi = 24\pi G a^2 ( \delta p_{\varphi} + \delta p_{\sigma} + \delta p_{B}), 
\label{two}\\
&& (h + 6 \xi)'' + 2 {\mathcal H} ( h + 6 \xi)' - 2 k^2 \xi = 24 \pi G a^2 \Pi_{\mathrm{B}},
\label{three}\\
&& k^2 \xi' = - 4\pi G [ \varphi' k^2 \chi_{\varphi} + \sigma' k^2 \chi_{\sigma} - P], 
\label{four}
\end{eqnarray}
where $ \chi_{\varphi}$ and $\chi_{\sigma}$ denote, respectively, the fluctuations 
of $\varphi$ and of $\sigma$.  Equations (\ref{one}) and (\ref{four}) come, respectively, from 
the $(00)$ and $(0i)$ components of the perturbed Einstein equations. Equations (\ref{two}) and (\ref{three}) 
are derived from $(i=j)$ and $(i\neq j)$ components of the perturbed Einstein equations. 
In Eqs. (\ref{three}) and (\ref{four}) the three-divergence 
of both sides of the equations has been taken. Finally, in Eq. (\ref{four}) $P(k,\tau)$ denotes the Fourier transform of the three-divergence  of the Poynting vector\footnote{While $P$ can be neglected in comparison with the other contributions of the momentum constraint, $P' = - 4  {\mathcal H} P + k^2  (\delta \rho_{\mathrm{B}} - 3 \Pi_{\mathrm{B}})/3$. Even if $P$ has been neglected in the final expressions, its derivative has been used whenever needed.}, i.e. $\vec{\nabla}\cdot[\vec{E}\times\vec{B}]/(4\pi a^4)$. The electric fields have been neglected since they are suppressed all along the inflationary phase.  

In the synchronous gauge description, the fluctuations of the energy density and pressure of $\varphi$ and $\sigma$ are:
\begin{eqnarray}
&&\delta\rho_{\varphi} = \frac{1}{a^2} \biggl[\varphi' \chi_{\varphi}' + \frac{\partial V}{\partial\varphi} a^2 \chi_{\varphi} \biggr], \quad 
\delta\rho_{\sigma} = \frac{1}{a^2} \biggl[\sigma' \chi_{\sigma}' + \frac{\partial W}{\partial\sigma} a^2 \chi_{\sigma}\biggr],
\label{deltarho}\\
&&\delta p_{\varphi} = \frac{1}{a^2} \biggl[\varphi' \chi_{\varphi}' - \frac{\partial V}{\partial\varphi} a^2 \chi_{\varphi}\biggr], \quad 
\delta p_{\sigma} = \frac{1}{a^2} \biggl[\sigma' \chi_{\sigma}' - \frac{\partial W}{\partial\varphi} a^2 \chi_{\sigma}\biggr].
\label{deltap}
\end{eqnarray}
Te fluctuations of the inflaton and of the spectator field obey, respectively, the following two equations:
\begin{eqnarray}
&& \chi_{\varphi}'' + 2 {\mathcal H} \chi_{\varphi}'  + k^2 \chi_{\varphi} + \frac{\partial^2 V}{\partial\varphi^2} a^2 \chi_{\varphi} - \frac{\varphi'}{2} h' =0,
\label{chi1}\\
&& \chi_{\sigma}'' + 2 {\mathcal H} \chi_{\sigma}'  + k^2 \chi_{\sigma} + \frac{\partial^2 W}{\partial\sigma^2} a^2 \chi_{\sigma} - \frac{\varphi'}{2} h' + \frac{\lambda_{\,, \sigma}}{\lambda}  \delta \rho_{\mathrm{B}} =0.
\label{chi2}
\end{eqnarray}
\subsection{Curvature perturbations}
The system of the perturbed Einstein equations (\ref{one})--(\ref{four}) supplemented by Eqs. (\ref{chi1}) and (\ref{chi2}) 
can be decoupled in terms of two variables defined as: 
\begin{equation}
q_{\varphi} = a \chi_{\varphi} - z_{\varphi} \xi, \qquad ,\qquad
q_{\sigma} = a \chi_{\sigma} - z_{\sigma} \xi, 
\label{qq}
\end{equation}
where $z_{\varphi} = a \varphi'/{\mathcal H}$ and $z_{\sigma} = a \sigma'/{\mathcal H}$.
The evolution equation obeyed by $q_{\varphi}$ and $q_{\sigma}$ are obtained from Eqs. (\ref{chi1}) and (\ref{chi2}) with the help of Eqs. (\ref{one}) and (\ref{four}).
The general result can be further simplified under the assumption that the energy density of $\sigma$ can be neglected in comparison 
with the energy density of the inflaton. Defining the rescaled Planck mass\footnote{In the present paper we shall use both $M_{\mathrm{P}}$ and $\overline{M}_{\mathrm{P}}$. The two 
quantities are equal up to the factor $\sqrt{8 \pi}$, i.e.   $\overline{M}_{\mathrm{P}} = M_{\mathrm{P}}/\sqrt{8\pi} = 1/\sqrt{8\pi G}$.} the Friedmann equations 
\begin{equation}
3 \overline{M}_{\mathrm{P}}^2\,{\mathcal H}^2 = \frac{1}{2}({\varphi'}^2 + {\sigma'}^2) +a^2\, V(\varphi) + a^2\,W(\sigma), \quad 
2 \overline{M}_{\mathrm{P}}^2\, ({\mathcal H}^2 - {\mathcal H}' )= {\varphi'}^2 + {\sigma'}^2,
\label{FLL}
\end{equation}
imply that the evolution equations for $q_{\varphi}$ and $q_{\sigma}$ can be expressed as 
\begin{eqnarray}
q_{\varphi}'' + k^2 q_{\varphi} - \frac{z_{\varphi}''}{z_{\varphi}} q_{\varphi} + \frac{a^3}{\overline{M}_{\mathrm{P}}}\, \overline{S}_{\varphi}(k,\tau) =0,
\label{prox7}\\
q_{\sigma}'' + k^2 q_{\sigma} - \frac{a''}{a} q_{\sigma} +  \frac{a^3}{\overline{M}_{\mathrm{P}}} \,\overline{S}_{\sigma}(k,\tau) =0.
\label{prox8}
\end{eqnarray}
Equations (\ref{prox7}) and (\ref{prox8}) hold under the approximation that $z_{\sigma} \ll z_{\varphi}$ (as implied by the occurrence that $\sigma' \ll \varphi'$); the source terms in Eqs. (\ref{prox7}) and (\ref{prox8}) are given by 
\begin{eqnarray}
&&\overline{{\mathcal S}}_{\varphi}(\vec{k},\tau) = \frac{\varphi'}{3 \overline{M}_{\mathrm{P}}\, {\mathcal H}}  \delta\rho_{\mathrm{B}}(\vec{k},\tau) + \frac{\varphi'}{2 \overline{M}_{\mathrm{P}}\, {\mathcal H}}\Pi_{\mathrm{B}}(\vec{k},\tau),
\label{prox5}\\
&&\overline{{\mathcal S}}_{\sigma}(\vec{k},\tau) = \biggl(\frac{ \sigma'}{3 \overline{M}_{\mathrm{P}}\, {\mathcal H}} + \overline{M}_{\mathrm{P}}  
\frac{\lambda_{,\,\sigma}}{\lambda} \biggr)
\delta\rho_{\mathrm{B}}(\vec{k},\tau) + \frac{\sigma'}{2 \overline{M}_{\mathrm{P}}\, {\mathcal H}}\Pi_{\mathrm{B}}(\vec{k},\tau).
\label{prox6}
\end{eqnarray}

The curvature perturbations on comoving orthogonal hypersurfaces are expressible, in the synchronous gauge, solely in terms of $\xi$ and of its first time derivative, i.e.
\begin{equation}
{\mathcal R} = \xi + \frac{{\mathcal H}\, \xi'}{{\mathcal H}^2 - {\mathcal H}'}.
\label{curv1}
\end{equation}
 Using then Eq. 
(\ref{four}) into Eq. (\ref{curv1}) to eliminate $\xi'$ and recalling Eq. (\ref{qq}),  ${\mathcal R}(\vec{k},\tau)$ becomes:
\begin{equation}
{\mathcal R}(\vec{k},\tau) \equiv - \frac{ z_{\varphi}(\tau)\,q_{\varphi}(\vec{k},\tau) + z_{\sigma}(\tau)\,q_{\sigma}(\vec{k},\tau)}{z_{\varphi}^2(\tau) + z_{\sigma}^2(\tau)} \simeq - \frac{q_{\varphi}(\vec{k},\tau)}{z_{\varphi}(\tau)} - q_{\sigma}(\vec{k},\tau) \frac{z_{\sigma}(\tau)}{z_{\varphi}^2(\tau)},
\label{curv2}
\end{equation}
where the second equality follows again in the limit $z_{\sigma} \ll z_{\varphi}$.
Equations (\ref{prox5}) and (\ref{prox6}) are easily solvable with standard Green's functions methods when the 
relevant wavelengths are larger than the Hubble radius during inflation; the result of this procedure can be written as 
\begin{eqnarray}
q_{\varphi}(\vec{k},\tau) &=& q_{\varphi}^{(1)}(\vec{k},\tau) - \overline{M}_{\mathrm{P}}\biggl[ c_{\varphi}\Omega_{\mathrm{B}}(\vec{k},\tau) + d_{\varphi} \Omega_{\mathrm{B}\Pi}(\vec{k},\tau) \biggr] a(\tau),
\label{prox16}\\
q_{\sigma}(\vec{k},\tau) &=& q_{\sigma}^{(1)}(\vec{k},\tau) - \overline{M}_{\mathrm{P}}\biggl[ c_{\sigma}\Omega_{\mathrm{B}}(\vec{k},\tau) + d_{\sigma} \Omega_{\mathrm{B}\Pi}(\vec{k},\tau) \biggr] a(\tau),
\label{prox17}
\end{eqnarray}
where the sources have been evaluated to leading order in $k\tau$ and 
\begin{equation}
\Omega_{\mathrm{B}}(\vec{k},\tau) = \frac{\delta\rho_{\mathrm{B}}(\vec{k},\tau)}{ 3 H^2 \overline{M}_{\mathrm{P}}^2}, \qquad \Omega_{\mathrm{B}\Pi}(\vec{k},\tau) = \frac{\Pi_{\mathrm{B}}(\vec{k},\tau)}{ 3 H^2 \overline{M}_{\mathrm{P}}^2}.
\label{prox18}
\end{equation}
In Eqs. (\ref{prox16}) and (\ref{prox17}) the following quantities have been introduced
\begin{eqnarray}
c_{\varphi} &=& \frac{m(f,\epsilon)}{3 \overline{M}_{\mathrm{P}}} \biggl(\frac{\varphi'}{{\mathcal H}} \biggr),\qquad d_{\varphi} = \frac{3}{2} c_{\varphi}
\label{prox17a}\\
c_{\sigma} &=& m(f,\epsilon)\,\biggl[\frac{1}{3 \overline{M}_{\mathrm{P}}} \biggl(\frac{\sigma'}{{\mathcal H}} \biggr) +\overline{M}_{\mathrm{P}}\frac{\lambda_{,\sigma}}{\lambda}\biggr],\qquad  d_{\sigma} = \frac{m(f,\epsilon)}{2 \overline{M}_{\mathrm{P}}}\,\biggl(\frac{\sigma'}{{\mathcal H}}\biggr),
\nonumber\\
m(f,\epsilon) &=& \frac{3 ( 1 - \epsilon)^2 }{( 1 - 2 f) ( 4 - 2 f - 3 \epsilon)}.
\label{prox17b}
\end{eqnarray}
Using Eq. (\ref{prox16}) and (\ref{prox17}) inside Eq. (\ref{curv2}) the resulting expression for the curvature perturbations can be written as:
\begin{equation}
{\mathcal R}(k,\tau) = {\mathcal R}_{*}(\vec{k}) + {\mathcal S}_{*}(\vec{k}) + {\mathcal M}^{(\mathrm{B})}_{\varphi\sigma}(\tau) \Omega_{\mathrm{B}}(k,\tau) + 
 {\mathcal M}^{(\mathrm{B}\Pi)}_{\varphi\sigma}(\tau) \Omega_{\mathrm{B}\Pi}(k,\tau),
\label{finex1}
\end{equation}
where ${\mathcal R}_{*}(\vec{k})$ denotes the standard adiabatic solution associated with $q^{(1)}_{\varphi}(\vec{k},\tau)$, ${\mathcal S}_{*}(\vec{k})$ is the non-adiabatic mode associated with $q^{(1)}_{\sigma}(\vec{k},\tau)$:
\begin{equation}
{\mathcal R}_{*}(\vec{k}) = - \frac{q^{(1)}_{\varphi}(\vec{k},\tau)}{z_{\varphi}(\tau)}, \qquad {\mathcal S}_{*}(\vec{k}) = - q^{(1)}_{\sigma}(\vec{k},\tau) \biggl(\frac{z_{\sigma}(\tau)}{z_{\varphi}^2(\tau)} \biggr).
\end{equation}
The two functions 
${\mathcal M}^{(\mathrm{B})}_{\varphi\sigma}(\tau)$ and $ {\mathcal M}^{(\mathrm{B}\Pi)}_{\varphi\sigma}(\tau)$ are defined as
\begin{eqnarray}
 {\mathcal M}^{(\mathrm{B})}_{\varphi\sigma}(\tau) \ =  \frac{\overline{M}_{\mathrm{P}}\, a(\tau)}{z_{\varphi}(\tau)} \biggl[ c_{\varphi} +  \biggl(\frac{z_{\sigma}}{z_{\varphi}}\biggr) \, c_{\sigma}\biggr],\quad 
 {\mathcal M}^{(\mathrm{B}\Pi)}_{\varphi\sigma}(\tau) =  \frac{\overline{M}_{\mathrm{P}}\, a(\tau)}{z_{\varphi}(\tau)} \biggl[ d_{\varphi} +  \biggl(\frac{z_{\sigma}}{z_{\varphi}}\biggr) \, d_{\sigma}\biggr].
\end{eqnarray}
The result obtained in Eq. (\ref{finex1}) has been deduced in rather general terms 
and it holds, with different forms of the coefficients, not only when the 
gauge fields are coupled to the spectator field but also when some direct coupling to the inflaton is present. In the forthcoming sections we shall assume the result of Eq. (\ref{finex1}) and keep the coefficients general. In terms of these coefficients the initial conditions of the Einstein-Boltzmann 
hierarchy can be derived. In the specific case discussed here the dominant contributions will be the ones of $c_{\varphi}$ and $d_{\varphi}$ which are, at most, ${\mathcal O}(1/\epsilon)$ in the nearly scale-invariant limit. 

\subsection{Global variables}
There is a conservation law associated with the evolution of ${\mathcal R}$. Recalling the explicit expression of Eq. (\ref{curv1}), we can take the first derivative 
of both sides and use the evolution equations of the fluctuations (\ref{one})--(\ref{four}) where, instead of 
specifying the energy densities and the pressures we simply introduce $\delta \rho_{\mathrm{t}}$ (the fluctuation of the total energy density), $\delta p_{\mathrm{t}}$ (the fluctuation of the total pressure) 
and the total non-adiabatic pressure fluctuation $\delta p_{\mathrm{nad}} = \delta p_{\mathrm{t}} - c_{\mathrm{st}}^2 \delta \rho_{\mathrm{t}}$ where $c_{\mathrm{st}}^2 = p_{\mathrm{t}}'/\rho_{\mathrm{t}}'$ is the total sound speed of the system; the background pressure and energy density $p_{\mathrm{t}}$ and $\rho_{\mathrm{t}}$ obey the conventional Friedmann equations in the spatially flat case: 
\begin{equation}
3 \overline{M}_{\mathrm{P}}^2\,{\mathcal H}^2 = a^2 \rho_{\mathrm{t}}
, \qquad 
2 \overline{M}_{\mathrm{P}}^2\, ({\mathcal H}^2 - {\mathcal H}' ) = a^2( \rho_{\mathrm{t}} +  p_{\mathrm{t}}).
\label{FLL1}
\end{equation}
Equation (\ref{FLL1}) implies $\rho_{\mathrm{t}}' + 3 {\mathcal H}(\rho_{\mathrm{t}} + p_{\mathrm{t}}) =0$.
The first derivative of ${\mathcal R}$ turns out to be 
\begin{eqnarray}
{\mathcal R}' = - \frac{{\mathcal H} \delta p_{\mathrm{nad}}}{p_{\mathrm{t}} + \rho_{\mathrm{t}}} + 
\frac{{\mathcal H}\delta \rho_{\mathrm{B}}}{p_{\mathrm{t}} + \rho_{\mathrm{t}}} \biggl( c_{\mathrm{st}}^2 - \frac{1}{3}\biggr)   + \frac{{\mathcal H}^2 (h' + 6 \xi')}{8\pi G a^2 ( p_{\mathrm{t}} + \rho_{\mathrm{t}})} - \frac{{\mathcal H} k^2 c_{\mathrm{st}}^2 \xi}{4\pi G a^2 (p_{\mathrm{t}} + \rho_{\mathrm{t}})} + 
\frac{{\mathcal H}\Pi_{\mathrm{t}}}{(p_{\mathrm{t}} + \rho_{\mathrm{t}})} .
\label{der4}
\end{eqnarray}
where  $\Pi_{\mathrm{t}}$ denote the total anisotropic stress of the system containing 
together with the anisotropic stress of the magnetic fields also the anisotropic stress of the other species. 
In the $\Lambda$CDM scenario the other source of anisotropic stress comes from the neutrino sector. 

If ${\mathcal S}_{*}(k) =0$ in Eq. (\ref{finex1}), $\delta p_{\mathrm{nad}} =0$ in Eq. 
(\ref{der4}). This will be the situation considered in the forthcoming section 
when discussing the magnetized adiabatic mode. Recalling the results of the first paper of Ref. \cite{magn3}, the presence of the 
non-adiabatic modes can also be easily considered along the lines illustrated above but for the purposes 
of the present discussion it is unnecessary. 

By taking the time derivative of both sides of Eq. (\ref{der4}) and from the evolution equations for $h$ and $\xi$
the equation of ${\mathcal R}$ simplifies even further:
\begin{equation}
{\mathcal R}'' + 2 \frac{z_{\mathrm{t}}'}{z_{\mathrm{t}}} {\mathcal R}' - c_{\mathrm{st}}^2 \nabla^2 {\mathcal R} = \Sigma_{{\mathcal R}}' + 2 \frac{z_{\mathrm{t}}'}{z_{\mathrm{t}}} \Sigma_{{\mathcal R}} + 
\frac{3 a^4}{z^2} \Pi_{\mathrm{t}},
\label{zt1}
\end{equation}
where 
\begin{eqnarray}
\Sigma_{{\mathcal R}} = - \frac{{\mathcal H}\, \delta p_{\mathrm{nad}}}{(p_{\mathrm{t}} + \rho_{\mathrm{t}})} + \frac{\mathcal H}{(p_{\mathrm{t}} + \rho_{\mathrm{t}})} \biggl[ \biggl(c_{\mathrm{st}}^2 - \frac{1}{3}\biggr) \delta\rho_{\mathrm{B}} + \Pi_{\mathrm{t}} \biggr],\qquad
z_{\mathrm{t}} = \frac{a^2 \sqrt{p_{\mathrm{t}} + \rho_{\mathrm{t}}}}{{\mathcal H} c_{\mathrm{st}}}.
\label{zt3}
\end{eqnarray}
The results of Eqs. (\ref{zt1})--(\ref{zt3}) reduce, in the absence of magnetic fields and in the absence of anisotropic stress, 
to the results of Lukash valid in the case of an irrotational relativistic fluid \cite{lukash,strokov,luk2}.  For wavelengths 
larger than the Hubble radius the Laplacian of ${\mathcal R}$ can be neglected and the solution of Eq. (\ref{zt1}) is
\begin{equation}
{\mathcal R}(\vec{x},\tau) = {\mathcal R}_{*}(\vec{x}) + \int_{\tau_{*}}^{\tau} d\tau' \Sigma_{\mathcal R}(\vec{x},\tau') + 
\int_{\tau_{*}}^{\tau} \frac{d\tau''}{z_{\mathrm{t}}^2(\tau'')} \int_{\tau_{*}}^{\tau''} \, a^4(\tau') \, \Pi_{\mathrm{t}}(\vec{x}, \tau') \, d\tau'.
\label{expans2}
\end{equation}
The continuity of ${\mathcal R}$ across the inflation-radiation transition can be verified in explicit toy models. 
Consider, for instance, the situation where $\rho_{\mathrm{t}}$ and $p_{\mathrm{t}}$ are both continuous across the 
inflation-radiation transition. Denoting with $\beta = a/a_{*}$ the normalized scale factor across the 
transition, we must require that the effective barotropic index $w_{\mathrm{t}} \to -1$ and $w_{\mathrm{t}} \to 1/3$  
when $\beta$ gets, respectively, much smaller and much larger than $1$.  An interesting interpolating 
solution of Friedmann equations with this property is given by
\begin{equation}
 \rho_{\mathrm{t}} = \frac{12 H_{*}^2 \, \overline{M}_{\mathrm{P}}^2}{(\beta^2 + 1)^2}, \qquad p_{\mathrm{t}} = 4 H_{*}^2 \overline{M}_{\mathrm{P}}^2 \, \frac{(\beta^2 - 3)}{(\beta^2 + 1)^3}.
 \label{int}
 \end{equation}
The barotropic index $w_{\mathrm{t}} = (1/3) (\beta^2 - 3)/(\beta^2 + 1)$ goes to $1/3$ for $\beta \gg 1 $ and to 
$-1$ for $\beta \ll 1$. In conformal time the evolution equations can be explicitly 
integrated with the result that  ${\mathcal H} = (\tau^2 + \tau_{*}^2)^{-1/2}$ and 
$\beta(\tau) = \biggl( \tau + \sqrt{\tau^2 + \tau_{*}^2}\biggr)$. These expressions can be used to verify the continuity of ${\mathcal R}(\vec{x},\tau)$ for instance, in Eq. (\ref{expans2}). 

When inflationary magnetic fields and  non-adiabatic pressure fluctuations are both vanishing, the initial conditions of the temperature and polarization anisotropies follow from Eqs. (\ref{zt1}), (\ref{zt3}) and (\ref{expans2}). Even if inflationary magnetic fields are 
absent the total anisotropic stress $\Pi_{\mathrm{t}}$ receives contribution from the neutrinos. The compatibility of the neutrino anisotropic 
stress with the other evolution equations implies, in the case of adiabatic initial conditions, the well known result \cite{cosm,cmbfast} stipulating that $\Pi_{\mathrm{t}} \simeq {\mathcal O}(k^2 \tau^2)$ when the relevant wavelengths are larger than the Hubble 
radius at the corresponding epoch. Using Eq. (\ref{expans2}) and the interpolating 
solution of Eq. (\ref{int}) it can be shown by direct integration that during the radiation epoch ${\mathcal R}(k,\tau) \simeq 
{\mathcal R}_{*}(k) \{ 1 - (4/9)[R_{\nu}/(4 R_{\nu} + 15)] k^2 \tau^2\}$ where $R_{\nu}<1$ is the (massless) neutrino fraction 
of the radiation plasma. In summary, for adiabatic initial conditions and it the absence of magnetic fields of inflationary origin 
${\mathcal R}(k,\tau) \simeq {\mathcal R}_{*}(k)$ with corrections which are always small for typical wavelengths larger 
than the Hubble radius. The generalization of this statement to the case when the inflationary seeds are present 
will be, among other things, the subject of the following section. 
\renewcommand{\theequation}{3.\arabic{equation}}
\setcounter{equation}{0}
\section{Initial conditions of the CMB anisotropies}
\label{sec3}
The value of curvature perturbations computed in the previous section will now be used as initial condition for 
the subsequent post-inflationary evolution. Focussing on the magnetized adiabatic mode the post-inflationary value of curvature perturbations will be parametrized as:
\begin{equation}
{\mathcal R}(k,\tau) ={\mathcal R}_{*}(\vec{k}) + c_{B} R_{\gamma} \Omega_{\mathrm{B}}(k) +  d_{B} R_{\gamma} \sigma_{\mathrm{B}}(k),
\label{in1}
\end{equation}
where $c_{B}$ and $d_{B}$ are of the same order and, as discussed before, at most ${\mathcal O}(1/\epsilon)$. In Eq. (\ref{in1})  the energy density and the anisotropic stress of the magnetic fields have been rescaled through the energy density of the photons. 
After neutrino decoupling (occurring approximately around the MeV) the anisotropic stress of the neutrinos is generated and $\Pi_{\mathrm{t}}(k,\tau)$ can be parametrized, in the $\Lambda$CDM scenario, as
\begin{equation}
\Pi_{\mathrm{t}}(k,\tau) = (p_{\nu} + \rho_{\nu}) \sigma_{\nu}(k,\tau) + (p_{\gamma} + \rho_{\gamma}) \sigma_{\mathrm{B}}(k),
\end{equation}
where, following the conventions of section \ref{sec2}, $\sigma_{\nu}$ is related to the anisotropic stress in real 
space as $\partial_{i}\partial_{j} \Pi^{ij}_{\nu} = (p_{\nu} + \rho_{\nu}) \nabla^2 \sigma_{\nu}$.

\subsection{Pre-decoupling plasma}
The perturbed Einstein equations for the post-inflationary system of magnetized perturbations include photons, neutrinos, baryons\footnote{The evolution of electrons and ions can be described, effectively, in terms 
of a single fluid called the baryon fluid whose velocity is the center of mass velocity of the electron-ion system. This happens because the Coulomb coupling is always tight, as argued long ago in the  pioneering work of Peebles and Yu \cite{peeblesyu}.} and CDM particles. The analog of Eqs. (\ref{one})--(\ref{four}) will then be:
\begin{eqnarray}
&& 2 k^2 \xi - {\mathcal H} h' = \frac{a^2}{\overline{M}_{\mathrm{P}}^2} ( \delta \rho_{\mathrm{t}} + \delta \rho_{\mathrm{B}}),
\label{HAM}\\
&& h'' + 2 {\mathcal H} h' - 2 k^2 \xi = \frac{ 3 a^2}{\overline{M}_{\mathrm{P}}^2}(\delta p_{\mathrm{t}} + \delta p_{\mathrm{B}}),
\label{heq}\\
&& ( h + 6 \xi)'' + 2 {\mathcal H} ( h + 6 \xi)' - 2 k^2 \xi =  \frac{ 3 a^2}{\overline{M}_{\mathrm{P}}^2} [(p_{\nu} + \rho_{\nu})\sigma_{\nu} + (p_{\gamma} + 
\rho_{\gamma}) \sigma_{\mathrm{B}}],
\label{anisstr}\\
&& k^2 \xi' = - \frac{a^2}{2\overline{M}_{\mathrm{P}}^2} (p_{\mathrm{t}} + \rho_{\mathrm{t}}) \theta_{\mathrm{t}},
\label{MOM}
\end{eqnarray}
where $\theta_{\mathrm{t}}$, $\delta\rho_{\mathrm{t}}$ and $\delta p_{\mathrm{t}}$ are\footnote{In Fourier space $\theta_{\mathrm{t}}$ and $\theta_{\mathrm{a}}$ denote, respectively, 
 the three-divergences of the total velocity field and of the different species composing the plasma.}:
\begin{equation}
(p_{\mathrm{t}} + \rho_{\mathrm{t}}) \theta_{\mathrm{t}} = \sum_{\mathrm{a}} (p_{\mathrm{a}} + \rho_{\mathrm{a}}) \theta_{\mathrm{a}},\qquad \delta \rho_{\mathrm{t}} = \sum_{\mathrm{a}} \delta_{\mathrm{s}} \rho_{\mathrm{a}}, \qquad
\delta p_{\mathrm{t}} = \sum_{\mathrm{a}} \delta_{\mathrm{s}} p_{\mathrm{a}} = w_{\mathrm{a}} \delta_{\mathrm{s}} \rho_{\mathrm{a}}.
\label{global}
\end{equation}
The sums appearing in Eq. (\ref{global}) extend over the four species of the plasma (i.e. photons, neutrinos, 
baryons and CDM particles) and $w_{\mathrm{a}}$ is the barotropic index of each species. By using the background equations in their general form, i.e. Eq. (\ref{FLL1}), Eqs. (\ref{HAM}) and  (\ref{MOM}) can also be written in more explicit terms as:
\begin{eqnarray}
&&2 k^2 \xi - {\mathcal H} h' = 3 {\mathcal H}^2\biggl\{ \Omega_{\mathrm{R}} (R_{\gamma} \delta_{\gamma} + 
R_{\nu} \delta_{\nu}) + R_{\gamma} \Omega_{\mathrm{R}} \Omega_{\mathrm{B}}  + 
 \Omega_{\mathrm{M}} \biggl[\biggl(\frac{\Omega_{\mathrm{c}0}}{\Omega_{\mathrm{M}0}}\biggr) \delta_{\mathrm{c}}
+ \biggl(\frac{\Omega_{\mathrm{b}0}}{\Omega_{\mathrm{M}0}}\biggr)\delta_{\mathrm{b}}\biggr]\biggr\},
\label{HAM1}\\
&& k^2 \xi' = - 2 {\mathcal H}^2 \biggl[ \Omega_{\mathrm{R}} R_{\nu}\theta_{\nu} +  \Omega_{\mathrm{R}} R_{\gamma}( 1 + R_{\mathrm{b}}) \theta_{\gamma\mathrm{b}} + \frac{3}{4} \Omega_{\mathrm{M}} \biggl(\frac{\Omega_{\mathrm{c}0}}{\Omega_{\mathrm{M}0}}\biggr)\rho_{\mathrm{c}} \theta_{\mathrm{c}} \biggr],
\label{MOM1}
\end{eqnarray}
where $\delta_{\gamma}$, $\delta_{\nu}$, $\delta_{\mathrm{b}}$ and $\delta_{\mathrm{c}}$ denote, with obvious notations, the density 
contrasts of the four species of the plasma. 

At sufficiently early times, the velocity of the photons coincides with the baryon velocity, i.e. $\theta_{\gamma} \simeq \theta_{\mathrm{b}} = \theta_{\gamma\mathrm{b}}$ since the photons and the baryons are 
tightly coupled by Thomson scattering. In the momentum constraint of Eq. (\ref{MOM1}) the tight-coupling 
regime has been already assumed. Furthermore, because we are going to solve the system
across the radiation-matter transition, the Poynting vector can be safely 
neglected. Using the same strategy leading to Eqs. (\ref{HAM1}) and (\ref{MOM1}), Eqs. (\ref{heq}) and (\ref{anisstr}) can be recast in the following form:
\begin{eqnarray}
&&  h'' + 2 {\mathcal H} h' - 2 k^2 \xi = 3 {\mathcal H}^2 \Omega_{\mathrm{R}} ( R_{\gamma} \delta_{\gamma} + R_{\nu} \delta_{\nu} +  R_{\gamma} \Omega_{\mathrm{B}})
\label{heq2}\\
&& ( h + 6 \xi)'' + 2 {\mathcal H} ( h + 6 \xi)' - 2 k^2 \xi = 12 {\mathcal H}^2 \Omega_{\mathrm{R}} [ R_{\nu} \sigma_{\nu} + R_{\gamma} \sigma_{\mathrm{B}}],
\label{anisstr2}
\end{eqnarray}
where, the critical fraction of radiation and of matter are defined as: 
\begin{equation}
\Omega_{\mathrm{R}} = \frac{a_{\mathrm{e}}}{a_{\mathrm{eq}} + a}, \qquad \Omega_{\mathrm{M}} = \frac{a}{a + a_{\mathrm{eq}}}.
\label{OMdef}
\end{equation}

The initial conditions of the Einstein-Boltzmann hierarchy are fully specified by solving also the equations of the monopoles and the dipoles of the phase-space distribution of the various species. The monopoles are related to the density contrasts whose evolution is given by:
\begin{eqnarray}
&& \delta_{\nu}' = - \frac{4}{3} \theta_{\nu} + \frac{2}{3} h',\qquad  \delta_{\gamma}' = - \frac{4}{3} \theta_{\gamma\mathrm{b}} + \frac{2}{3} h',
\label{deltanugamma}\\
&& \delta_{\mathrm{b}}' = -\theta_{\gamma\mathrm{b}} + \frac{h'}{2},\qquad 
\delta_{\mathrm{c}}' = - \theta_{\mathrm{c}} + \frac{h'}{2}.
\label{deltabc}
\end{eqnarray}
The  equations for the dipoles are related to the peculiar velocity of the various species: 
\begin{eqnarray}
&&\theta_{\nu}' = \frac{k^2}{4} \delta_{\nu} - k^2 \sigma_{\nu},
\label{thetanu}\\
&&\theta_{\gamma\mathrm{b}}'+ \frac{{\mathcal H} R_{\mathrm{b}}}{R_{\mathrm{b}} +1 } \theta_{\gamma\mathrm{b}} = \frac{k^2}{4 ( 1 + R_{\mathrm{b}})} \delta_{\gamma} + \frac{k^2}{4 ( 1 + R_{\mathrm{b}})} (\Omega_{\mathrm{B}} - 4 \sigma_{\mathrm{B}}),
\label{thetab}\\
&& \theta_{\mathrm{c}}' + {\mathcal H} \theta_{\mathrm{c}} =0.
\label{thetac}
\end{eqnarray}
In Eq. (\ref{thetanu}) there appears also $\sigma_{\nu}$ whose 
evolution equation is coupled to all the higher multipoles of the neutrino phase-space distribution
\begin{eqnarray}
\sigma_{\nu}' &=& \frac{4}{15} \theta_{\nu} - \frac{3}{10} k {\mathcal F}_{\nu 3} - \frac{2}{15} h' - \frac{4}{5} \xi',
\label{sigmanu}\\
{\mathcal F}_{\nu\ell}' &=& \frac{k}{2\ell + 1} \biggl[ \ell {\mathcal F}_{\nu(\ell -1)} - (\ell + 1) {\mathcal F}_{\nu(\ell+1)}\biggr],\qquad \ell\geq 3.
\label{Fell}
\end{eqnarray}
where ${\mathcal F}_{\nu\ell}$ denotes the $\ell$th multipole of the perturbed phase-space distribution of the neutrinos.
The evolution equations reported here hold prior to photon decoupling. Across decoupling the baryons and the 
photons obey effectively different equations since the approximation based on the tight photon-baryon coupling breaks down. 

\subsection{Explicit solutions and magnetized adiabatic mode}
The solution of Eq. (\ref{FLL1}) across the radiation-matter transition and for a spatially flat Universe reads:
\begin{equation}
\alpha = \frac{a}{a_{\mathrm{eq}}}= x^2 + 2 x, \qquad x = \frac{\tau}{\tau_{1}}, \qquad \tau_{1} = \frac{2}{H_{0}} \sqrt{\frac{a_{\mathrm{eq}}}{\Omega_{\mathrm{M}  0}}}  \simeq 283.73 \,\, \biggl(\frac{ h_{0}^2 \Omega_{\mathrm{M}0}}{0.1368}\biggr)^{-1}\, \mathrm{Mpc},
\label{alpha}
\end{equation}
where $a_{\mathrm{eq}}$ is the scale factor at the equality already introduced in Eq. (\ref{OMdef}), i.e. the 
moment when non-relativistic matter and radiation contribute equally to the total energy density of the plasma.
For  $\alpha = \rho_{\mathrm{M}}/ \rho_{\mathrm{R}} \ll 1$ (i.e. $a \ll a_{\mathrm{eq}}$) the plasma is dominated by radiation and according to Eq. (\ref{alpha}),  $\alpha \simeq 2 x  + {\mathcal O}(x^2)= 2 (\tau/\tau_{1})$. 

Defining as $\tau_{\mathrm{i}}$ the initial integration time,
 it will be required that $k \tau_{\mathrm{i}} < 1$ for all the modes involved in the calculations. 
The double expansion employed in setting initial conditions
of the truncated Einstein-Boltzmann hierarchy can be formally expressed as\footnote{In terms of the solution of Eq. (\ref{alpha}), 
${\mathcal H} = 2 \sqrt{\alpha + 1}/(\tau_{1} \alpha)$ and $c_{\mathrm{st}}^2 = 4/( 3 \alpha + 4)$.}
\begin{equation}
\alpha \ll 1, \qquad \frac{k}{ a H} = \frac{k}{ {\mathcal H}} = \frac{\kappa \,\alpha}{2\,\sqrt{\alpha + 1}} \simeq 
k \tau \ll 1.
\label{expans1}
\end{equation}
where $\kappa = k \tau_{1}$ measures how large the wavelength was, in Hubble units, around equality (note, indeed, that $\tau_{\mathrm{eq}} = (\sqrt{2} -1) \tau_{1} \simeq \tau_{1}/2$).

The initial conditions studied here incorporate the inflationary seeds in the adiabatic mode and belong to the class of the magnetized adiabatic modes (see \cite{magn2,magn3} and references therein). To investigate the time evolution of the system it is useful to employ directly the normalized scale factor $\alpha$. The initial conditions pertaining to the magnetized adiabatic mode can then be written as:
\begin{eqnarray}
&& \delta_{\nu}(\kappa,\alpha_{\mathrm{i}}) \simeq \delta_{\gamma}(\kappa,\alpha_{\mathrm{i}}) \simeq \frac{3}{4} \delta_{\mathrm{b}}(\kappa,\alpha_{\mathrm{i}}) \simeq \frac{3}{4} \delta_{\mathrm{c}} = - R_{\gamma}\Omega_{\mathrm{B}}(\kappa,\alpha_{\mathrm{i}}),
\label{IN1}\\
&& \theta_{\nu}(\kappa,\alpha_{\mathrm{i}}) \simeq \theta_{\gamma \mathrm{b}}(\kappa,\alpha_{\mathrm{i}}) \simeq \theta_{\mathrm{c}}(\kappa,\alpha_{\mathrm{i}}) \simeq 0,
\label{IN2}\\
&& \sigma_{\nu}(\kappa, \alpha_{\mathrm{i}}) =0,\qquad {\mathcal F}_{\ell} (\kappa,\alpha_{\mathrm{i}}) =  0,
\label{IN3}
\end{eqnarray}
where $\alpha_{\mathrm{i}}$ denotes the initial integration variable and $\ell \geq 3$. 
In the synchronous gauge, the density contrasts on uniform curvature hypersurfaces are defined as
$\zeta_{\mathrm{a}} = \xi + \delta_{\mathrm{a}}/[3(w_{\mathrm{a}} + 1)]$. But then, the conditions (\ref{IN1}) imply that all all the $\zeta_{\mathrm{a}}$ must be equal, i.e. 
$\zeta_{\nu} = \zeta_{\gamma} =\zeta_{\mathrm{c}}= \zeta_{\mathrm{b}}$  for $\alpha = \alpha_{\mathrm{i}}$ and $\kappa < 1$. In this sense the initial conditions (\ref{IN1})--(\ref{IN3}) generalize the adiabatic mode 
to the situation where inflationary seeds are present\footnote{It should be borne in mind that, 
in the synchronous gauge, the condition $\theta_{\mathrm{c}}=0$ is enforced not so much because of a property of the initial 
data but rather to fix completely the coordinate system and to avoid the occurrence of known spurious (gauge) modes arising 
in the synchronous description \cite{syn}}.
 
According to Eq. (\ref{zt1}) the initial conditions (\ref{IN1})--(\ref{IN3}) seem to be compatible with the adiabatic mode only if $\Pi_{\mathrm{t}} \to 0$. The neutrino anisotropic 
stress must be zero initially and also its first derivative is zero  since ${\mathcal F}_{\nu\ell} =0$ for $\ell \geq 3$. But thanks to the presence of the magnetic anisotropic stress, the total anisotropic stress does not vanish even before neutrino decoupling: $\Pi_{\mathrm{t}}(\kappa,\alpha_{\mathrm{i}}) \neq 0$ 
even if  $\sigma_{\nu}(\kappa, \alpha_{\mathrm{i}})$ and its derivatives are all vanishing. 
Direct numerical integration shows that, after a transient time,  $\Pi_{\mathrm{t}} \to 0$ even if, initially, $\Pi_{\mathrm{t}} \neq 0$.
This result is established by integrating Eqs. (\ref{heq})--(\ref{anisstr2}), (\ref{deltanugamma})--(\ref{deltabc}) and (\ref{thetanu})--(\ref{Fell}) in the background defined by Eq. (\ref{alpha}) and subjected to the initial conditions (\ref{in1}) and (\ref{IN1})--(\ref{IN3}).
\begin{figure}[!ht]
\centering
\includegraphics[height=6cm]{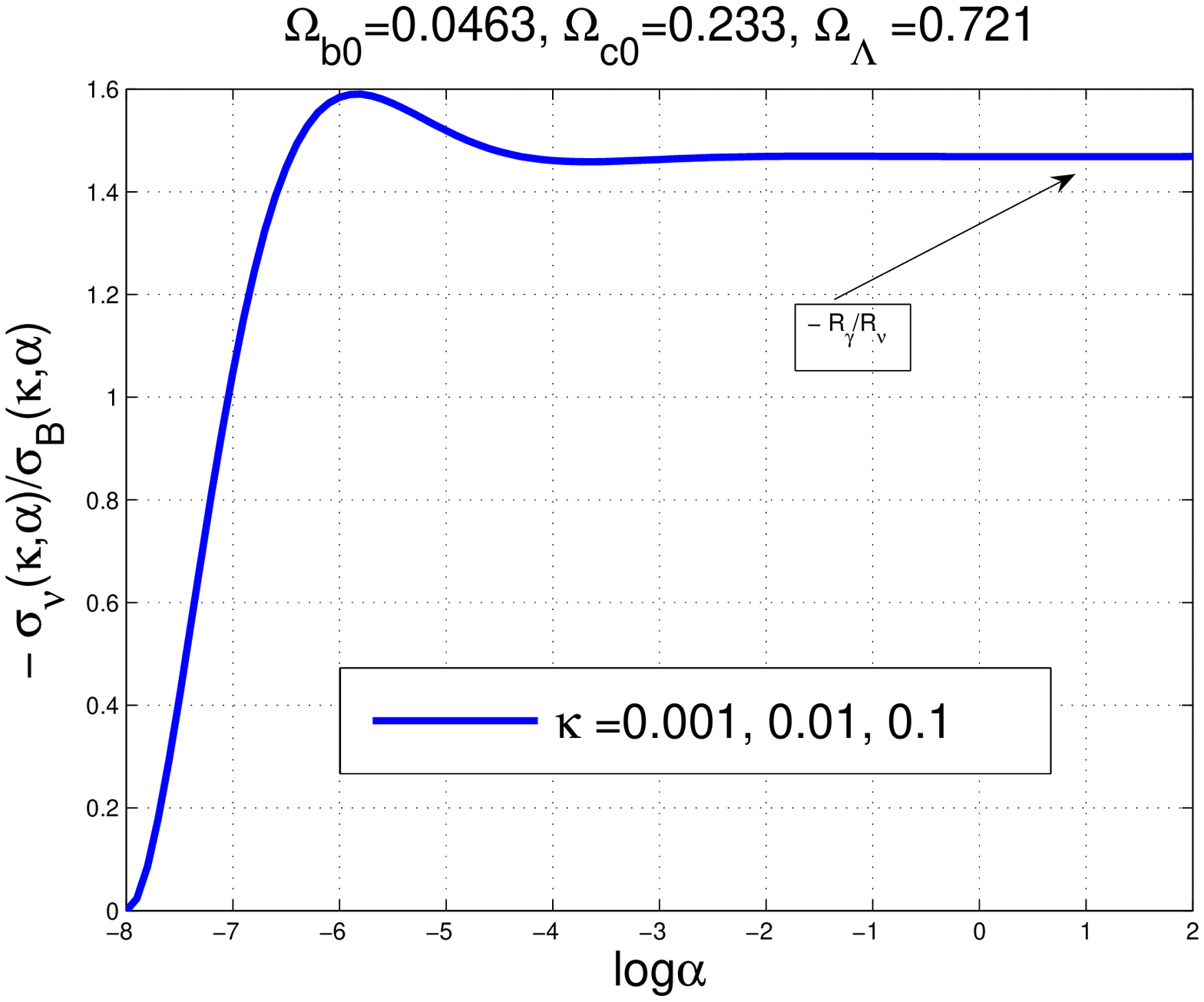}
\includegraphics[height=6cm]{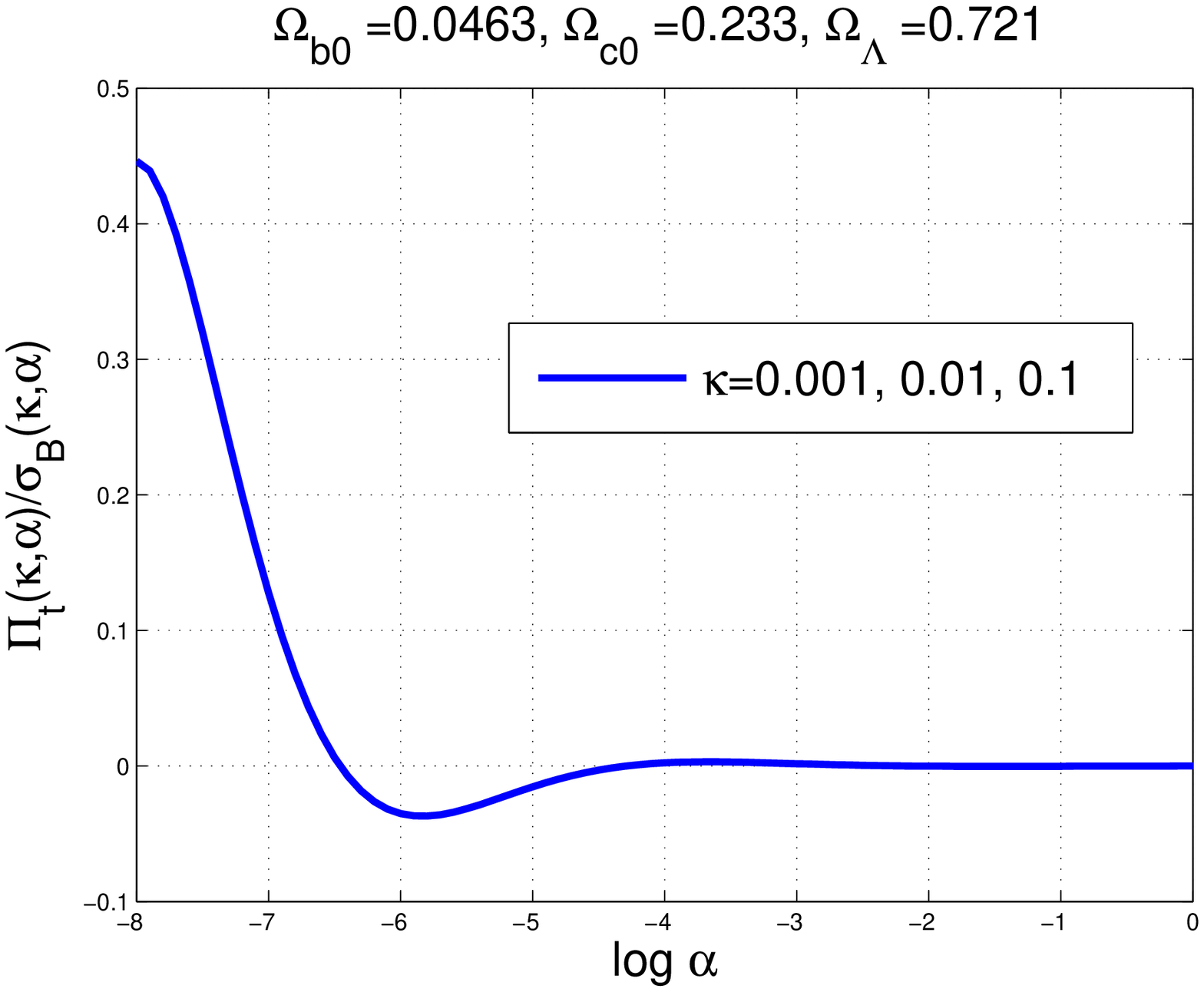}
\caption[a]{Evolution of the anisotropic stress in the case of the magnetized adiabatic mode. In the left plot 
the neutrino anisotropic stress is illustrated. In the plot at the right the total anisotropic stress is reported. 
On the horizontal axes the results are given in terms of the common logarithm of the normalized scale factor.}
\label{Figure1}      
\end{figure}
In Fig. \ref{Figure1} the evolution of the neutrino anisotropic $\sigma_{\nu}(\kappa,\alpha)$ stress and of the total anisotropic stress 
$\Pi_{\mathrm{t}}(\kappa,\alpha)$ is illustrated as a function of the common logarithm of the normalized scale 
factor. On the vertical axis of both plots of Fig. \ref{Figure1} the anisotropic stresses are given in units 
of the magnetic anisotropic stress $\sigma_{\mathrm{B}}(\kappa)$.

Denoting with $R_{\nu}$ and $R_{\gamma} = 1 - R_{\nu}$  the fractions of neutrinos and photons in the pre-decoupling plasma\footnote{In the $\Lambda$CDM paradigm 
$R_{\nu} = [3 \times(7/8)\times (4/11)^{4/3}]/[1 + 3 \times(7/8)\times (4/11)^{4/3}]=0.4052$ where $3$ counts the massless neutrino families, $(7/8)$ stems from the Fermi-Dirac statistics and $(4/11)^{4/3}$ is related to the different kinetic temperature of neutrinos.},
the arrow in the left plot underlines the value $-R_{\gamma}/R_{\nu}$ (which is asymptote of $\sigma_{\nu}/\sigma_{\mathrm{B}}$ for $k\tau \ll 1$).  The result of Fig. \ref{Figure1} can be obtained by integrating directly the system in the $\alpha$ parametrization. The equations are reported, for completeness, in appendix \ref{APPB}. As the legends indicate in Fig. \ref{Figure1}, different values of $\kappa$ produce results which 
are indistinguishable provided $\kappa \ll 1$. 

In the limit  $\alpha < 1$ and $k \tau < 1$ we can solve consistently the system discussed in the previous section and expressed, in the $\alpha$-parametrization, in appendix \ref{APPB}. The result is:
\begin{eqnarray}
\xi(k,\tau) &=& {\mathcal R}_{*}(k)\biggl[ 1 - \frac{( 4 R_{\nu} + 5) k^2 \tau^2}{12 ( 4 R_{\nu} + 15)}\biggr] + 
R_{\gamma} \Omega_{\mathrm{B}}(k)\biggl[ c_{B} - \frac{R_{\nu} k^2 \tau^2}{6 ( 4 R_{\nu} + 15)}\biggr]
\nonumber\\
&+& R_{\gamma}\sigma_{B} \biggl[ d_{B} + \frac{2 k^2 \tau^2}{3 (4 R_{\nu} + 15)}\biggr]
\label{xi1}\\
h(k,\tau) &=& \frac{{\mathcal R}_{*}(k)}{2} k^2 \tau^2\biggl[ 1 + \frac{8 R_{\nu}^2 - 14 R_{\nu} - 75}{36(2 R_{\nu} + 25)(4 R_{\nu} + 15)} k^2 \tau^2\biggr] 
\nonumber\\
&+& 
\frac{R_{\gamma} \Omega_{\mathrm{B}}(k)}{2} k^2 \tau^2 \biggl[ c_{B} + \frac{R_{\nu} ( 20 R_{\nu} - 15)}{180 ( 4 R_{\nu} +15) (2 R_{\nu} + 25)} k^2 \tau^2\biggr]
\nonumber\\
&+& \frac{R_{\gamma} \sigma_{\mathrm{B}}(k)}{2} k^2 \tau^2 \biggl[ d_{B} - \frac{R_{\nu} ( 20 R_{\nu} - 15)}{45 ( 4 R_{\nu} +15) (2 R_{\nu} + 25)} k^2 \tau^2\biggr].
\label{h1}
\end{eqnarray}
For the density contrasts we have instead
\begin{eqnarray}
\delta_{\gamma}(k,\tau) &=& \frac{{\mathcal R}_{*}(k)}{3} k^2 \tau^2 - R_{\gamma} \Omega_{\mathrm{B}}(k)\biggl[ 1 - 
\frac{1}{3} \biggl( c_{B} - \frac{R_{\nu}}{2 R_{\gamma}}\biggr) k^2 \tau^2 \biggr]
\nonumber\\
&+&  \frac{2}{3}\sigma_{\mathrm{B}}(k) \biggl[ 1 + \frac{d_{B} R_{\gamma}}{2} \biggr] k^2 \tau^2
\label{dg1}\\
\delta_{\nu}(k,\tau) &=& \frac{{\mathcal R}_{*}(k)}{3} k^2 \tau^2 - R_{\gamma} \Omega_{\mathrm{B}}(k)\biggl[ 1 + 
\biggl( R_{\gamma} - 2 c_{B}\biggr) \frac{k^2 \tau^2}{6} \biggr]
\nonumber\\
&+&  \frac{R_{\gamma}}{3}\sigma_{\mathrm{B}}(k) \biggl[ d_{B} - \frac{3}{2 R_{\nu}}\biggr] k^2 \tau^2
\label{dn1}\\
\delta_{\mathrm{c}}(k,\tau) &=& \frac{{\mathcal R}_{*}(k)}{4} k^2 \tau^2 -  \frac{3}{4} R_{\gamma} \Omega_{\mathrm{B}}(k) \biggl[ 1 - c_{B} \frac{k^2\tau^2}{3} \biggr] + \frac{d_{B} R_{\gamma}}{4} \sigma_{\mathrm{B}}(k) k^2 \tau^2 
\label{dc1}\\
\delta_{\mathrm{c}}(k,\tau) &=& \frac{{\mathcal R}_{*}(k)}{4} k^2 \tau^2 - \frac{3}{4} R_{\gamma} \Omega_{\mathrm{B}}(k) \biggl[ 1 - \frac{1}{4} \biggl(c_{B} - \frac{R_{\nu}}{2 R_{\gamma}} \biggr) k^2 \tau^2 \biggr] 
\nonumber\\
&+& \frac{\sigma_{\mathrm{B}}(k)}{4} k^2 \tau^2 ( 2 + d_{B} R_{\gamma}) 
\label{db1}
\end{eqnarray}
For the velocities we have
\begin{eqnarray}
\theta_{\gamma\mathrm{b}}(k,\tau) &=& \frac{k^4 \tau^3}{36} {\mathcal R}_{*}(k) +
\frac{R_{\gamma} \Omega_{\mathrm{B}}(k)}{4} k^2 \tau \biggl[ 1 + \frac{1}{9} \biggl(c_{B} \frac{R_{\gamma}}{R_{\nu}} - \frac{1}{2} \biggr) k^2 \tau^2 \biggr] 
\nonumber\\
&-& \sigma_{\mathrm{B}}(k) \biggl[ 1 +\frac{1}{18} \biggl( 1 - \frac{d_{B} R_{\gamma}}{2}\biggr) k^2\tau^2 \biggr]
\label{vgb1}\\
\theta_{\nu}(k,\tau) &=& {\mathcal R}_{*}(k) \biggl(\frac{4 R_{\nu} + 23}{4 R_{\nu} +15}\biggr) \frac{k^4 \tau^3}{36} 
- \frac{R_{\gamma} \Omega_{\mathrm{B}}(k)}{4} \biggl[ 1 - \frac{1}{9}\biggl(\frac{4 R_{\nu} + 23}{4 R_{\nu} +15} c_{B} - 
\frac{4 R_{\nu} + 27}{2(4 R_{\nu} + 15)} \biggr) k^2 \tau^2 \biggr]
\nonumber\\
&+& \sigma_{\mathrm{B}}(k) \frac{R_{\gamma}}{R_{\nu}} k^2 \tau \biggl\{ 1 + \frac{k^2\tau^2}{18} \biggl[ 
\frac{4 R_{\nu} + 27}{2(4 R_{\nu} + 15)} + \frac{d_{B} R_{\nu}}{2} \biggl(\frac{4 R_{\nu} + 23}{4 R_{\nu} +15}\biggr)\biggr]\biggr\}
\label{vnu1}
\end{eqnarray}
while $\theta_{\mathrm{c}} =0$. Finally the anisotropic stress of the neutrinos is given by
\begin{eqnarray}
\sigma_{\nu}(k,\tau) &=& - \frac{R_{\gamma}}{R_{\nu}} \sigma_{\mathrm{B}}(k) - \frac{2 {\mathcal R}_{*}(k)}{3 ( 4 R_{\nu} + 15)} k^2 \tau^2 
\nonumber\\
&-& \biggl[ \frac{2}{3( 4R_{\nu} +15)}\biggl( c_{B} + \frac{3}{4} \biggr) R_{\gamma} \Omega_{\mathrm{B}}(k) - 
\frac{2 \sigma_{\mathrm{B}}(k)}{(4 R_{\nu} + 15)} \frac{R_{\gamma}}{R_{\nu}} \biggl(1 - \frac{d_{B}}{3} R_{\nu}\biggr) \biggr]k^2 \tau^2.
\label{sn1}
\end{eqnarray}
All the higher multipoles in the neutrino hierarchy have been consistently set to zero since we are 
concerned here with the adiabatic initial data. Different initial conditions may  demand different 
assumptions on the higher multipoles of the hierarchy. This analysis closely follows earlier 
results (see \cite{magn2,magn3} and references therein) where, for the first time, the problem 
of the magnetized initial conditions of the Einstein-Boltzmann hierarchy has been posed and discussed. 
The crucial difference is represented by the inflationary origin of the 
magnetic fields which induces a further contribution on the curvature perturbations and, therefore, on the 
whole hierarchy.  Note that when the magnetized contribution is switched off, the initial conditions of Eqs. 
(\ref{xi1})--(\ref{sn1}) reproduce the standard adiabatic initial condition discussed long ago (see, e.g. 
\cite{cosm}).
\renewcommand{\theequation}{4.\arabic{equation}}
\setcounter{equation}{0}
\section{Temperature and polarization power spectra}
\label{sec4}
The temperature and polarization observables can be obtained by integrating numerically the magnetized Einstein-Boltzmann hierarchy across recombination with the full set of initial conditions discussed from Eq. (\ref{xi1}) to Eq. (\ref{sn1}). The Boltzmann 
integrator employed here is based on the code described in Ref. \cite{magn2,magn3} and used to investigate the magnetized CMB anisotropies. 
The Boltzmann code is based on Cosmics \cite{cosm} and on CMBFAST \cite{cmbfast} and it includes 
the evolution of magnetic fields within a consistent magnetohydrodynamical approximation. 

The temperature autocorrelations (TT correlations in what follows), the polarization 
autocorrelations (EE correlations in what follows) and the cross-correlation between the temperature and the polarization (TE correlations in what follows) are defined in the standard standard way (see, for instance, Eqs. (2.61)--(2.63) of  the second paper quoted in Ref. \cite{magn3}). 
All the different correlation spectra will be expressed in units of $(\mu \mathrm{K})^2$.  The following shorthand notation shall be used:
\begin{equation}
 {\mathcal G}^{(\mathrm{TT})}_{\ell} =  \frac{\ell (\ell  +1)}{2 \pi} C_{\ell}^{\mathrm{TT}}, \qquad 
 {\mathcal G}^{(\mathrm{EE})}_{\ell} =  \frac{\ell (\ell  +1)}{2 \pi} C_{\ell}^{\mathrm{EE}},\qquad  {\mathcal G}^{(\mathrm{TE})}_{\ell} =  \frac{\ell (\ell  +1)}{2 \pi} C_{\ell}^{\mathrm{TE}}.
\label{CORR1}
\end{equation}
In the minimal $\Lambda$CDM scenario the angular power spectra of Eq. (\ref{CORR1}) are the only non-vanishing observables since the tensor modes are neglected and the B-mode polarization is absent. As we shall argue later, however, a B-mode autocorrelation can be induced, via Faraday effect, from the EE correlations. 

The spectra of Eq. (\ref{CORR1}) depend on $6$ independent parameters 
\begin{equation}
{\mathcal G}_{\ell}^{(\mathrm{XY})}= {\mathcal G}_{\ell}^{(\mathrm{XY})}(n_{\mathrm{s}}, \,\Omega_{\mathrm{b}0}, \, \Omega_{\mathrm{c}0}, \Omega_{\Lambda},\, H_{0}, \epsilon_{\mathrm{re}}),
\label{CORR2}
\end{equation}
where $\epsilon_{\mathrm{re}}$ (not to be confused with the slow-roll parameter) denotes the optical depth at reionization.
\begin{figure}[!ht]
\centering
\includegraphics[height=6cm]{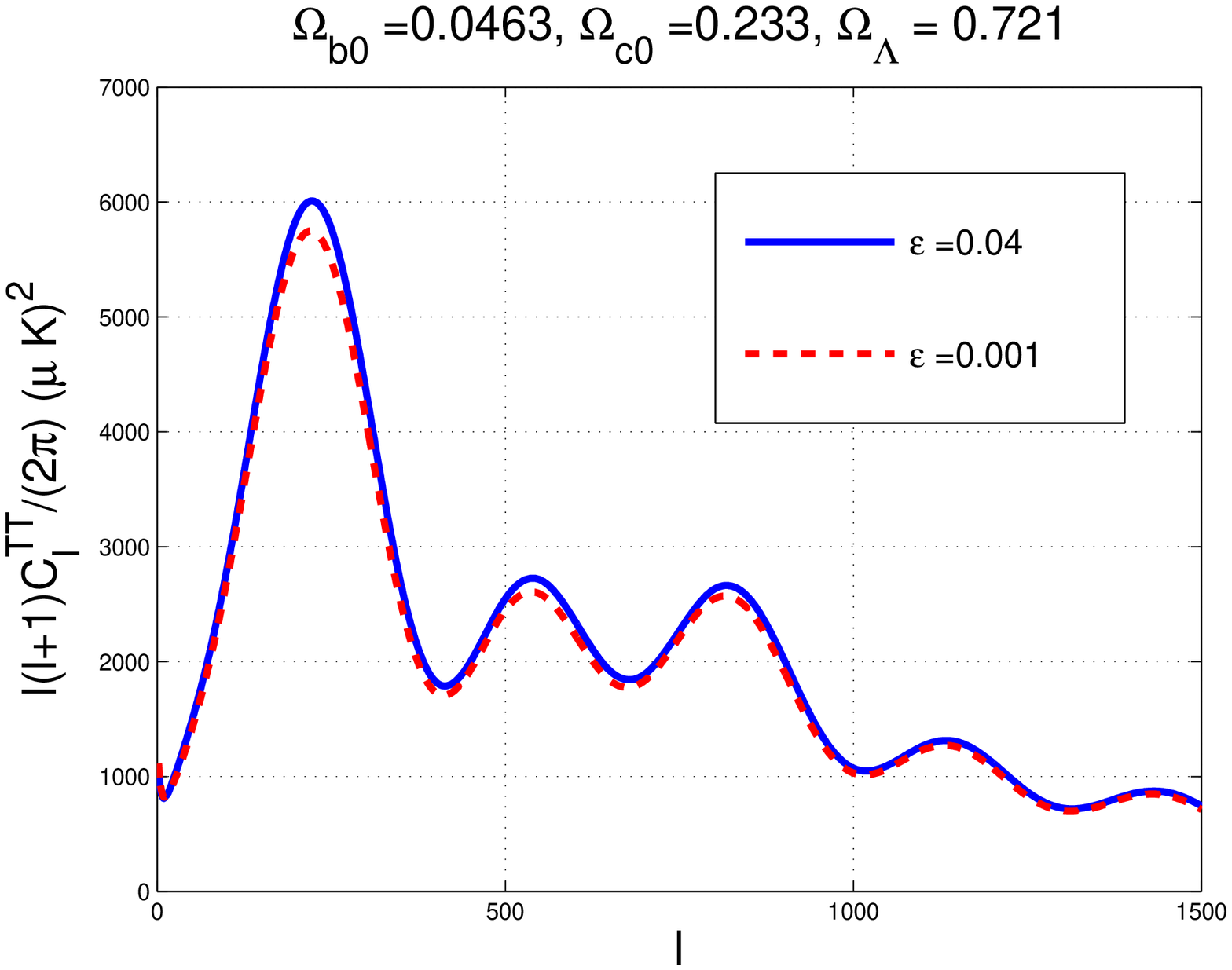}
\includegraphics[height=6cm]{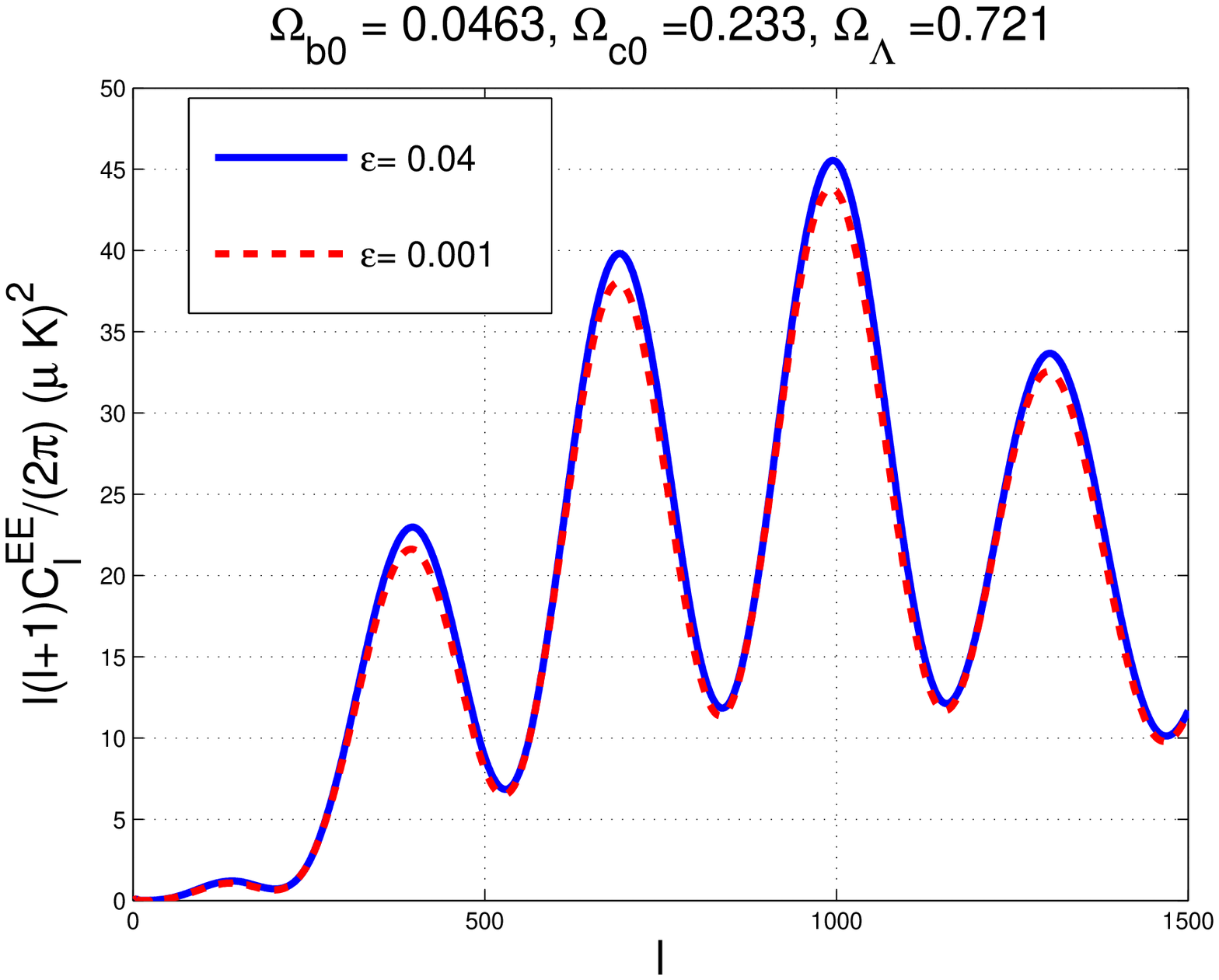}
\includegraphics[height=5.5cm]{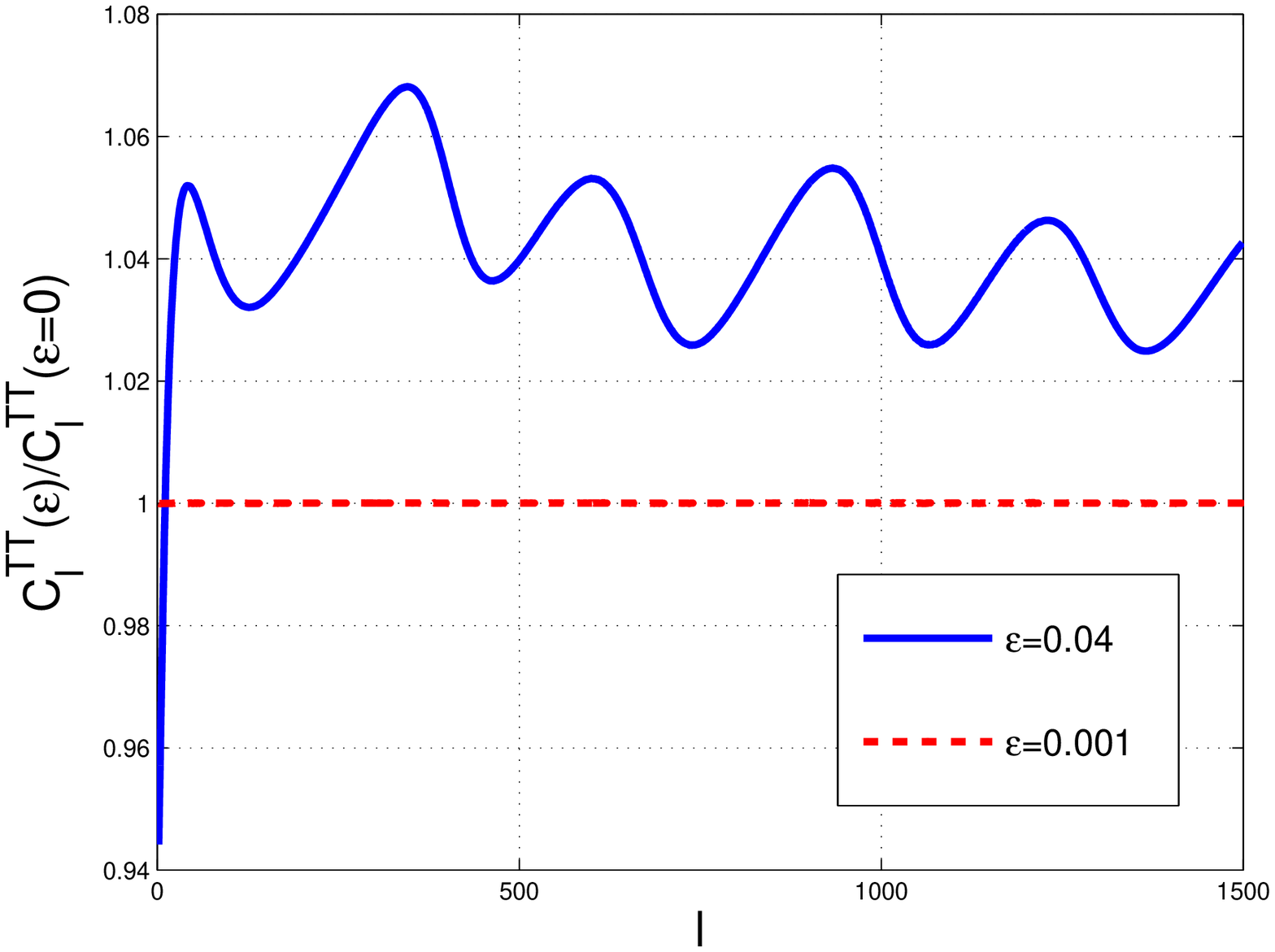}
\includegraphics[height=5.5cm]{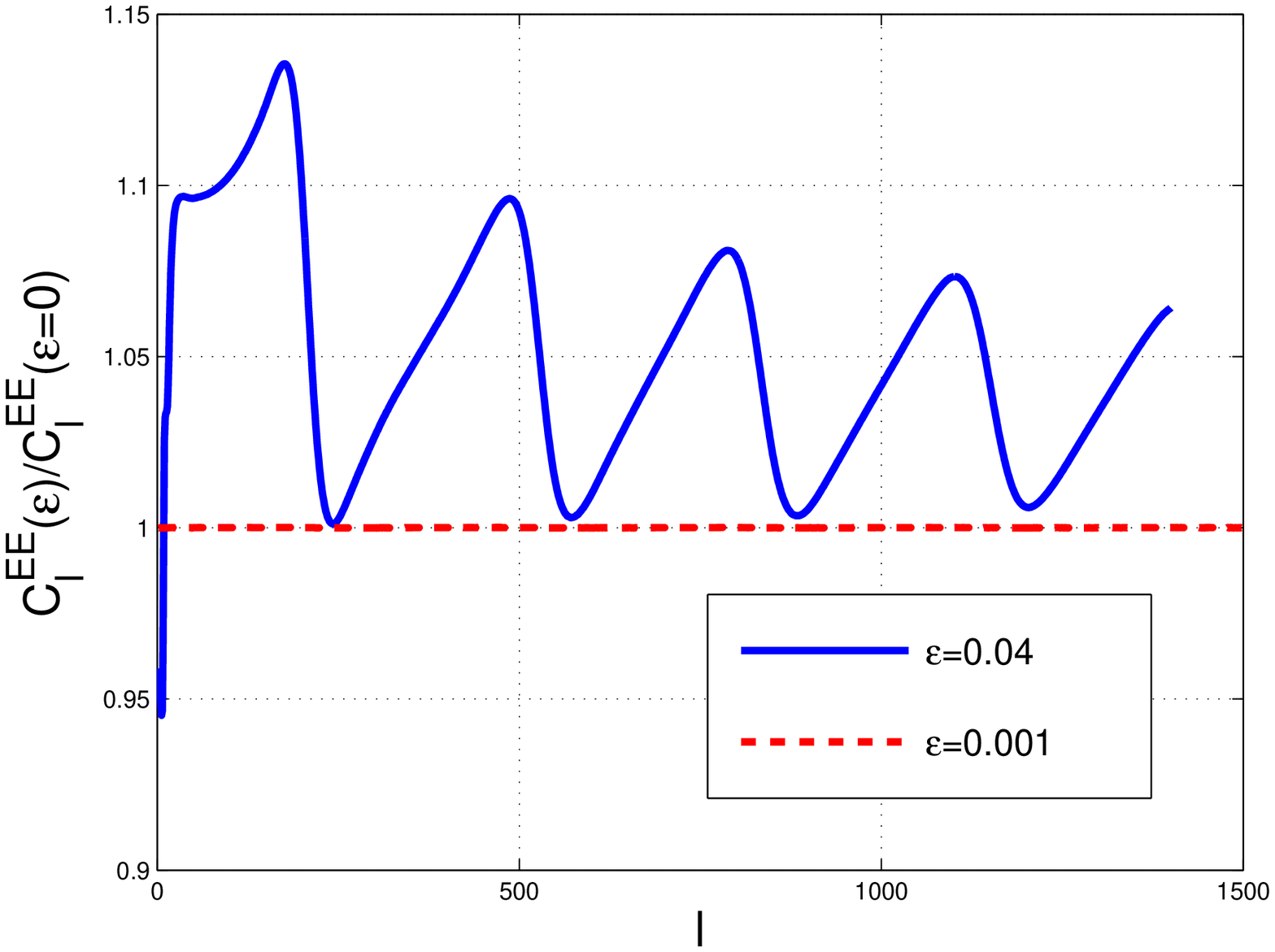}
\caption[a]{In the two upper plots the TT and the EE correlations are illustrated for the initial conditions 
of Eqs. (\ref{in1})--(\ref{sn1}) }
\label{Figure2}      
\end{figure}
The pivotal parameters of the $\Lambda$CDM paradigm can be determined on the basis of different data sets 
and, for illustrative purposes, we shall bound the attention only to three best fits. The first one is obtained by 
comparing the $\Lambda$CDM paradigm to the WMAP 9yr data alone (see, in particular, \cite{wmap1}):
\begin{equation}
( \Omega_{\mathrm{b}0}, \, \Omega_{\mathrm{c}0}, \Omega_{\mathrm{de}0},\, h_{0},\,n_{\mathrm{s}},\, \epsilon_{\mathrm{re}}) \equiv 
(0.0463,\, 0.233,\, 0.721,\,0.700,\, 0.972,\,0.089),
\label{ppp1}
\end{equation}
with ${\mathcal A}_{{\mathcal R}} = 2.41\times 10^{-9}$ (recall, in fact, the parametrization 
introduced in Eq. (\ref{av1a})). If we include the data sets pertaining to the baryon acoustic oscillations (see, e.g. \cite{SDSS}) the parameters are slightly different:
\begin{equation}
( \Omega_{\mathrm{b}0}, \, \Omega_{\mathrm{c}0}, \Omega_{\mathrm{de}0},\, h_{0},\,n_{\mathrm{s}},\, \epsilon_{\mathrm{re}}) \equiv 
(0.0477,\, 0.247,\, 0.705,\,0.686,\, 0.967,\,0.086),
\label{ppp2}
\end{equation}
with  ${\mathcal A}_{{\mathcal R}} = 2.35\times 10^{-9}$.
Another possible set of parameters considered hereunder is the one obtained by combining the WMAP9 data with 
the direct determinations of the Hubble rate
\begin{equation}
( \Omega_{\mathrm{b}0}, \, \Omega_{\mathrm{c}0}, \Omega_{\mathrm{de}0},\, h_{0},\,n_{\mathrm{s}},\, \epsilon_{\mathrm{re}}) \equiv 
(0.0445,\, 0.216,\, 0.740,\,0.717,\, 0.980,\,0.092),
\label{ppp3}
\end{equation}
with  ${\mathcal A}_{{\mathcal R}} = 2.45\times 10^{-9}$. 

The presence of inflationary magnetic fields affects
the CMB observables obtained in the framework of a particular best fit to the WMAP 9yr data for a sufficiently 
large value of $\epsilon$. 
\begin{figure}[!ht]
\centering
\includegraphics[height=6cm]{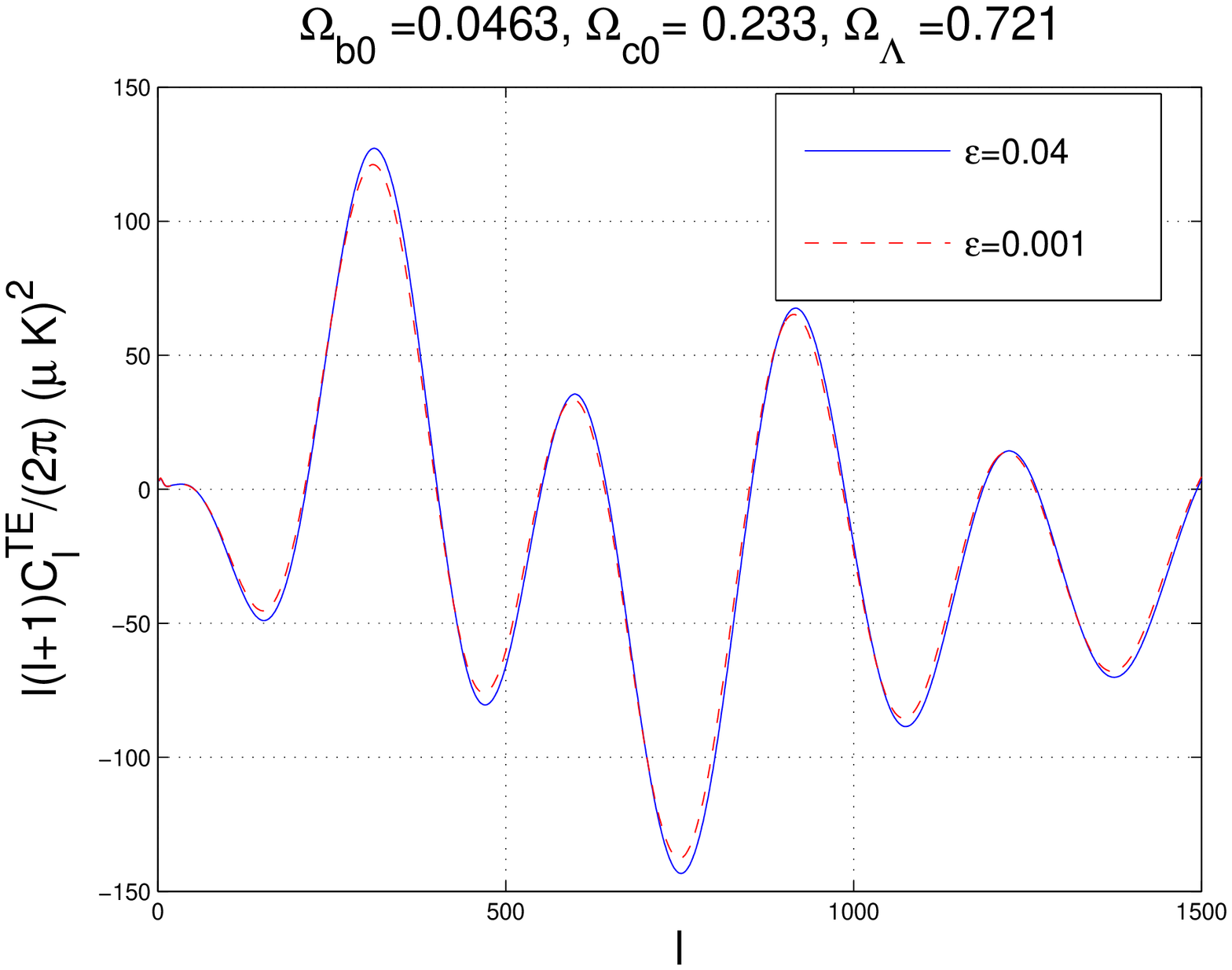}
\includegraphics[height=6cm]{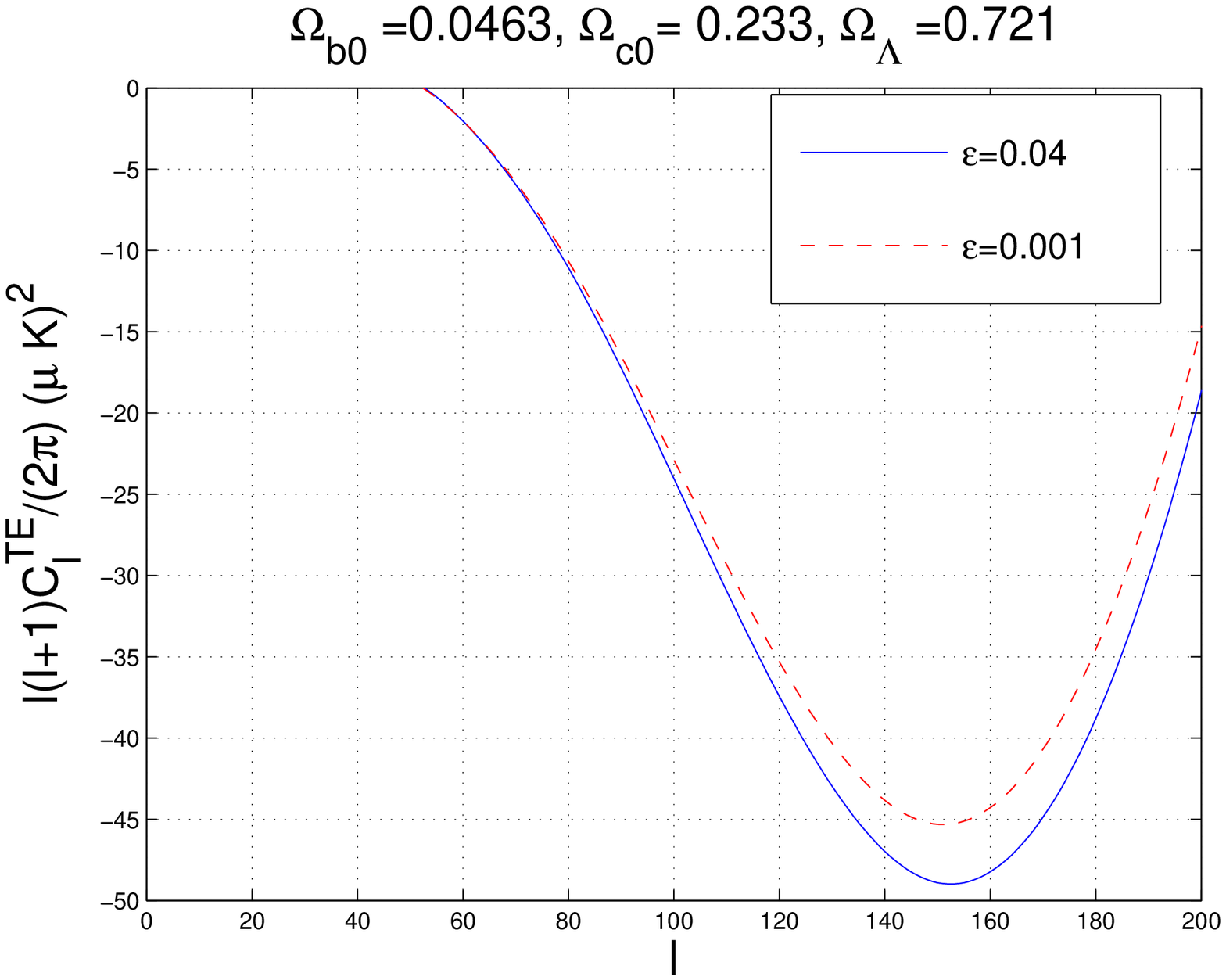}
\caption[a]{The same cases of Fig. \ref{Figure2} are illustrated in terms of the cross-correlation between 
temperature and E-mode polarization. At the right the detail of the first 
anticorrelation peak is illustrated.}
\label{Figure3}      
\end{figure}
This aspect is illustrated in Figs. \ref{Figure2} and \ref{Figure3} where  the magnetic fields have been introduced both at the level of the initial 
conditions and at the level of the evolution equations, as discussed in section \ref{sec3}. 
Both in Figs. \ref{Figure2} and \ref{Figure3} the fiducial set of $\Lambda$CDM parameters has been chosen as
in Eq. (\ref{ppp1}). In the framework of the $\Lambda$CDM scenario with standard thermal history the only 
extra parameter is represented by the slow-roll parameter $\epsilon$: the other 
parameters controlling the amplitude of the magnetic power spectrum of Eq. (\ref{sp1a}) 
are determined by the other parameters of the concordance scenario.

In Fig. \ref{Figure2} for sake of simplicity two extreme examples are illustrated, i.e. the case $\epsilon=0.04$ (full line) and $\epsilon=0.001$ (dashed line). In the two upper plots of Fig. \ref{Figure2} the TT and EE power spectra are reported, while in the two lower plots of Fig. \ref{Figure2}  the angular power spectra appearing in the upper plots have been divided by the best fit to the vanilla $\Lambda$CDM model (i.e. in the absence of inflationary magnetic fields). Using the notations of Eq. (\ref{CORR1}) in the two lower plots of Fig. \ref{Figure2} 
we illustrated, respectively, ${\mathcal G}_{\ell}^{(\mathrm{TT})}(\epsilon)/{\mathcal G}_{\ell}^{(\mathrm{TT})}(\epsilon=0)$ and ${\mathcal G}_{\ell}^{(\mathrm{EE})}(\epsilon)/{\mathcal G}_{\ell}^{(\mathrm{EE})}(\epsilon=0)$.
The shorthand notation $\epsilon=0$ simply means that the corresponding power spectrum is taken to be 
independent of $\epsilon$, as it happens in the vanilla $\Lambda$CDM with no tensors. In Fig. \ref{Figure3} the same analysis has been 
performed in the case of the TE correlations which cannot be simply divided 
by the corresponding WMAP 9yr best fit since the TE correlations are not positive definite. To illustrate 
more closely the differences between the different models, in the right plot of Fig. \ref{Figure3} the 
first anticorrelation peak has been shown in greater detail.

From Figs. \ref{Figure2} and \ref{Figure3} the case $\epsilon =0.001$ is practically indistinguishable from the WMAP 9yr bestfit while 
the case $\epsilon =0.04$ shows quantitive and qualitative differences potentially jeopardizing the agreement of the computed spectra with the observational data. This disagreement arises since the first, second and third  peaks of the acoustic oscillations 
are distorted. This aspect can be scrutinized from the values of the TT correlations in the neighborhood of the first three acoustic peaks. 
The position of the peaks will be denoted, respectively,  by $\ell_{1}$, $\ell_{2}$ and $\ell_{3}$. To pin down the position of the peaks in terms of the $\Lambda$CDM parameters we use the following parametrization adapted to WMAP 9yr data \cite{magn3}:
\begin{equation}
\ell_{j} = \overline{\ell}_{j} + \Delta\ell_{j}, \qquad \overline{\ell}_{j} = \ell_{A}( j - \phi_{j}), 
\label{prma1}
\end{equation}
where $\phi_{j}$ and $\Delta\ell_{j}$ are given, for $j=1,\,2,\,3$, as follows:
\begin{eqnarray}
&& \phi_{1} =  0.267 \, \biggl(\frac{r_{\mathrm{R}*}}{0.3} \biggr)^{0.1},\quad  \phi_{2} =  0.241\, \biggl(\frac{r_{\mathrm{R}*}}{0.3} \biggr)^{0.1},\quad   \phi_{3} =  0.353\, \biggl(\frac{r_{\mathrm{R}*}}{0.3} \biggr)^{0.1},
\label{prma2}\\
&& \Delta \ell_{1} = 0.13 \, |n_{\mathrm{s}} -1| \overline{\ell}_{1}, \qquad \Delta \ell_{2} = 0.33 \, |n_{\mathrm{s}} -1|\overline{\ell}_{2},\qquad \Delta \ell_{3} = 0.61 \, |n_{\mathrm{s}} -1| \overline{\ell}_{3}.
\label{prma3}
\end{eqnarray}
Note that $\ell_{A}$ is simply the well known acoustic multipole which is expressible through the angular diameter distance to recombination. Its standard expression can be reduced to a more explicit formula 
\begin{equation}
\ell_{\mathrm{A}} = \biggl(\frac{z_{*}}{10^{3}}\biggr)^{1/2} \frac{\sqrt{R_{\mathrm{b}*}}\,d_{\mathrm{A}}(z_{*})}{\ln{\biggl[
\frac{\sqrt{1 + R_{\mathrm{b}*}} + \sqrt{( 1 +  r_{\mathrm{R}*})R_{\mathrm{b}*}}}{1 + 
\sqrt{r_{\mathrm{R}*} R_{\mathrm{b}*}}}\biggr]}}, 
\label{prma4}
\end{equation}
where $ r_{\mathrm{R}*}$ and $R_{\mathrm{b}*}$ are given by 
\begin{equation}
r_{\mathrm{R}*} =\frac{\rho_{\mathrm{R}}(z_{*})}{\rho_{\mathrm{M}}(z_{*})} =
4.15 \times 10^{-2} \, (h_{0}^2 \Omega_{\mathrm{M}})^{-1}\, \biggl(\frac{z_{*}}{10^{3}}\biggr), \quad R_{\mathrm{b}}(z) = \frac{3}{4} \frac{\rho_{\mathrm{b}}}{\rho_{\gamma}} = 
30.36 \,h_{0}^2\Omega_{\mathrm{b}0} \,\biggl(\frac{10^{3}}{z_{*}}\biggr).
\label{prma5}
\end{equation}
The quantity $z_{*}$ is the redshift to recombination which can be directly expressed in terms 
of $\Lambda$CDM parameters as 
\begin{eqnarray}
z_{*} &=& 1048[ 1 + (1.24 \times 10^{-3})\, (h_{0}^2\Omega_{\mathrm{b}0})^{- 0.738}] [ 1 + g_{1} (h_{0}^2 \Omega_{\mathrm{M}0})^{\,\,\,g_2}],
\label{TP6}\\
g_{1} &=& \frac{0.0783 \,(h_{0}^2 \Omega_{\mathrm{b}})^{-0.238}}{[1 + 39.5 \,\,
(h_{0}^2 \Omega_{\mathrm{b}0})^{\,\,0.763}]},\qquad 
g_{2} = \frac{0.560}{1 + 21.1 \, (h_{0}^2 \Omega_{\mathrm{b}0})^{\,\,1.81}}.
\label{TP7}
\end{eqnarray}
The parameters of Eq. (\ref{ppp1}) imply 
$z_{*} =1090.95$ in excellent agreement with the estimate of  Ref.
\cite{wmap1,wmap2,wmap3}  (i.e. $z_{*} = 1090.41 \pm 0.57$) in the case of the WMAP 9yr data alone in the light of the vanilla $\Lambda$CDM scenario. The relative heights of the acoustic peaks computed 
 in the case of the best-fit of Eq. (\ref{ppp1}) are:
\begin{equation}
\overline{H}_{1} = \frac{{\mathcal G}^{(\mathrm{TT})}_{\ell_{1}}}{{\mathcal G}^{(\mathrm{TT})}_{\ell =10}} = 6.942,\qquad \overline{H}_{2} = \frac{{\mathcal G}^{(\mathrm{TT})}_{\ell_{2}}}{{\mathcal G}^{(\mathrm{TT})}_{\ell_{1}}} = 0.447, \qquad 
\overline{H}_{3} = \frac{{\mathcal G}^{(\mathrm{TT})}_{\ell_{3}}}{{\mathcal G}^{(\mathrm{TT})}_{\ell_{2}}} = 0.981,
\label{PS8}
\end{equation}
where $\ell_{1} = 221$,  $\ell_{2} = 538$ and  $\ell_{3} = 815$ are, respectively, the locations 
of the first three acoustic peaks obtainable from Eqs. (\ref{prma1})--(\ref{prma4}) and coincide, approximately, with the ones directly obtainable from the angular power spectra. In the case of the angular power spectra illustrated with the full line in the left plots of Fig. \ref{Figure2} the same ratios of Eq. (\ref{PS8}) are
$\overline{H}_{1}= 7.35$, $\overline{H}_{2} = 0.453$ and $\overline{H}_{3} = 0.976$. This shows 
that the largest effect is on the first peak while the other two are comparatively less affected. 

The results obtained in the case of the parameters of Eq. (\ref{ppp1}) are quantitatively and qualitatively 
close to the results obtainable in the cases of Eqs. (\ref{ppp2}) or (\ref{ppp3}).
\begin{figure}[!ht]
\centering
\includegraphics[height=6cm]{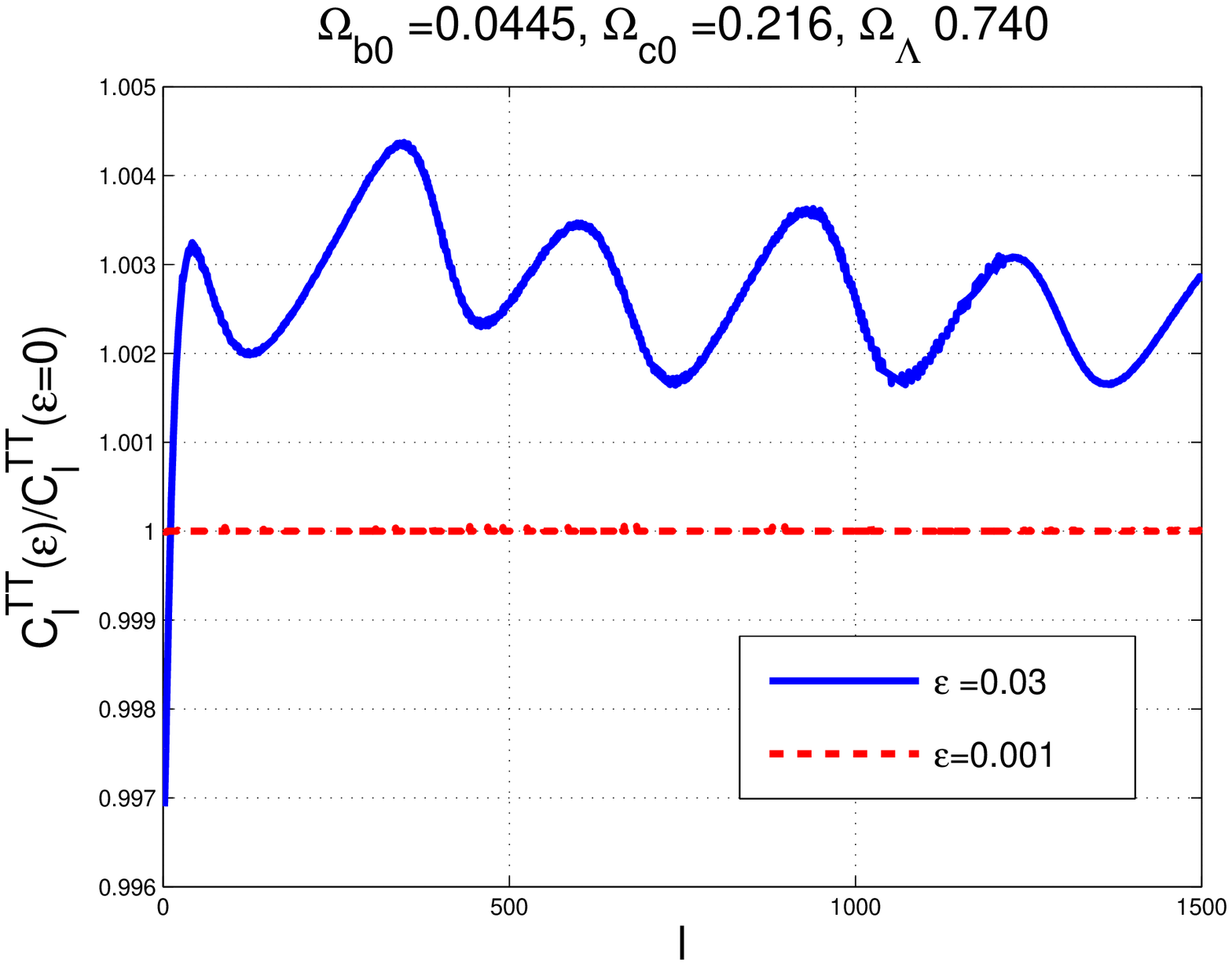}
\includegraphics[height=6cm]{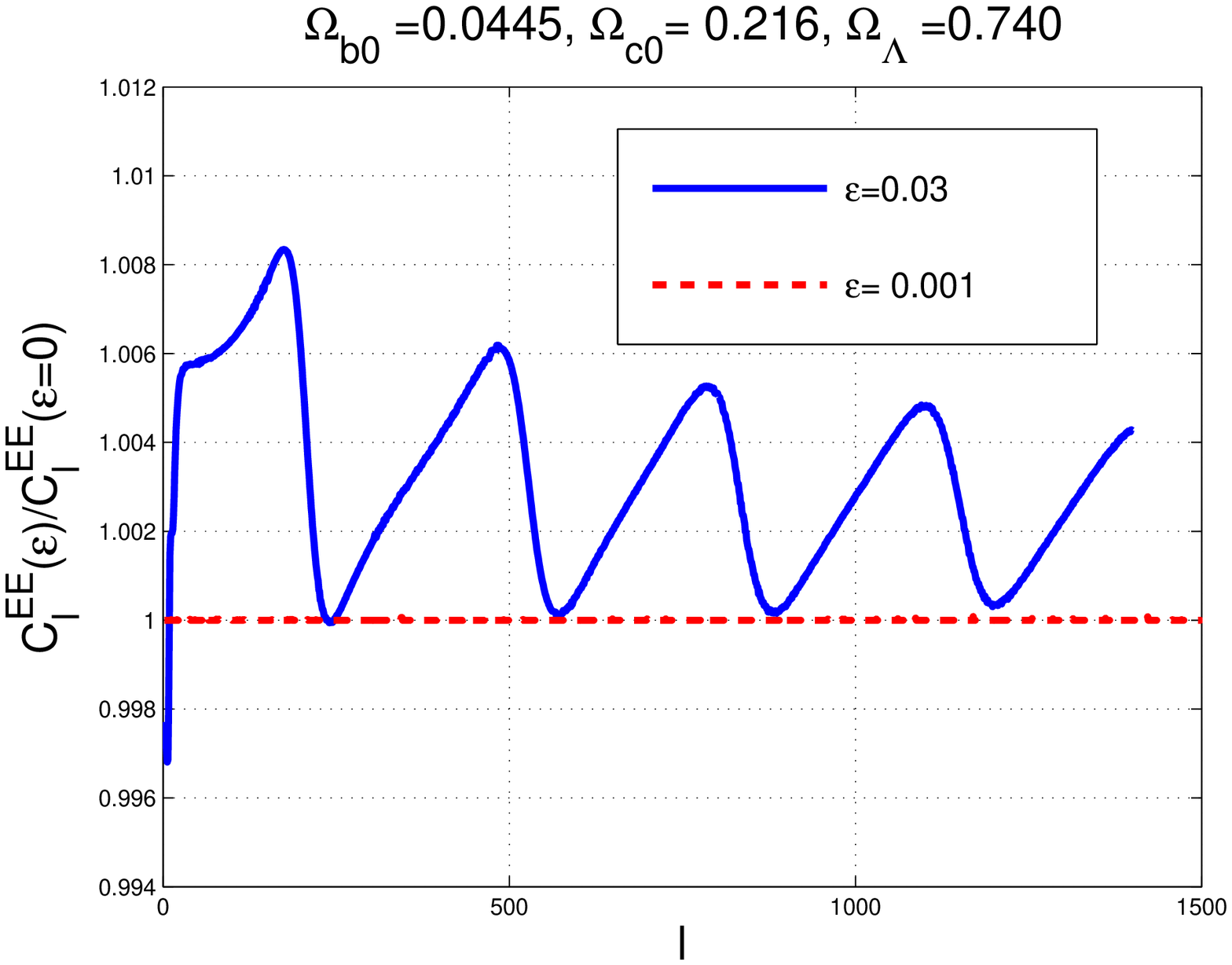}
\caption[a]{The temperature and polarization autocorrelations divided by the corresponding 
best fits are illustrated for the fiducial set of parameters reported in Eq. (\ref{ppp3}). }
\label{Figure4}      
\end{figure}
In Fig. \ref{Figure4}, for instance, the parameters of Eq. (\ref{ppp3}) have been assumed while 
one of the values of $\epsilon$ is different. From a closer comparison of  Figs. \ref{Figure2} and \ref{Figure4},
it can be argued that for $\epsilon<0.03$ the difference between the magnetized angular power and the corresponding $\Lambda$CDM best fit is smaller than $10^{-3}$.

In the $\Lambda$CDM paradigm without tensors there are no sources of B-mode polarization but a stochastic magnetic field can rotate the polarization plane of the CMB at a rate depending on the difference between the refractive indices associated, respectively, with the positive and with the negative 
helicities. Faraday rotation is one of the situations where the inadequacy of the one-fluid approximation (for the baryon-lepton fluid) is manifest. The positive and negative helicities composing the (linear) CMB polarization experience,
in a background magnetic field, two different phase velocities, two different dielectric contants and, ultimately, two different refractive indices. The mismatch between the refractive index of the positive and negative helicities  induces, effectively, a rotation of the CMB polarization and, hence, a B-mode. The inclusion of the Faraday effect in the treatment implies, physically, that the proton-electron fluid (sometimes dubbed as baryon fluid) should be treated as effectively composed by two different species, i.e. the electrons and the ions. 

The Faraday effect for the CMB polarization can be treated either with uniform magnetic fields or with stochastic magnetic fields and different analyses have been performed starting with the ones of Ref. \cite{far1} (see also \cite{far2} for an incomplete list of references). 
\begin{figure}[!ht]
\centering
\includegraphics[height=7cm]{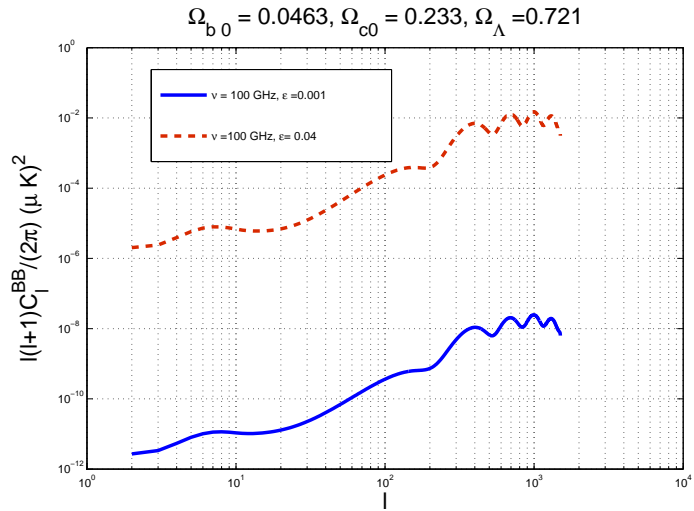}
\caption[a]{The B-mode polarization power spectrum induced by the Faraday rotation of the E-mode polarization.}
\label{Figure5}      
\end{figure}
Since the inflationary magnetic fields does not break spatial isotropy, the angular power 
spectrum of Faraday rotation can be written as:
\begin{equation}
\langle {\mathcal I}(\hat{n}_{1}) {\mathcal I}(\hat{n}_{2})\rangle = \frac{1}{4\pi}\sum_{\ell} (2 \ell + 1)C_{\ell}^{(\mathrm{F})} P_{\ell}(\hat{n}_{1} \cdot\hat{n}_{2}), \qquad {\mathcal I}(\hat{n}) = \frac{ 3}{16 \pi^2 e} \frac{\hat{n} \cdot \vec{B}}{\nu^2}.  
\label{Far1}
\end{equation}
where ${\mathcal I}(\hat{n})$ is a normalized form of the Faraday rotation rate, $\nu$ denotes the comoving 
frequency and $P_{\ell}(z)$ are the standard Legendre polynomials. In terms of the power spectrum of the Faraday rate the autocorrelation of the B-mode polarization 
can be computed as 
\begin{equation}
{\mathcal C}_{\ell}^{\mathrm{BB}} = \sum_{\ell_{1},\,\,\ell_{2}}  {\mathcal Z}(\ell, \ell_1, \ell_{2}) C_{\ell_{2}}^{\mathrm{EE}} C_{\ell_{1}}^{(\mathrm{F})}
\label{F14a}
\end{equation}
where ${\mathcal Z}(\ell, \ell_1, \ell_{2})$ is a complicated function of the multipole moments containing 
also a Clebsch-Gordon coefficient\footnote{For an explicit expression of ${\mathcal Z}(\ell, \ell_1, \ell_{2})$ see, for instance, the discussion contained in Appendix C of the last paper quoted in Ref. \cite{far2}.} while $C_{\ell_{2}}^{\mathrm{EE}}$ is the E-mode power spectrum 
already discussed above.  

In Fig. \ref{Figure5} the B-mode autocorrelation is computed in the case of the fiducial set of parameters 
of Eq. (\ref{ppp1}) and for two illustrative values of $\epsilon$. The Faraday rotation rate depends also on the 
frequency of observation which has been taken to be $100$ GHz. 
We remind that WMAP experiment observes the microwave sky in five frequency channels (i.e. $23$, $33$, $41$, $61$ and $94$ in units of GHz). The bandwidth increases from small to high frequencies signaling that probably the best sensitivity to polarization comes the high frequency channels.
The WMAP 9yr data do not report any direct detection of the B-mode polarization. 

The Planck experiment \cite{planck} is observing the microwave sky in nine frequency channels:
three frequency channels 
(i.e. $\nu= 30,\,44,\,70$ GHz) belong to the low frequency instrument (LFI);  six 
channels (i.e. $\nu= 100,\,143,\,217,\,353,\,545,\,857$ GHz) belong to the high 
frequency instrument (HFI). The BB power spectra for all the relevant 
frequency channels. There are reasons to expect that the sensitivity to polarization 
will be larger at high frequency \cite{planck}. At the same time the expected signal will be larger at small frequencies. We will illustrate the present results for a putative frequency of $100$ GHz.

The B-mode polarization induced by the tensor modes of the geometry is independent on the frequency channel 
and for $r_{\mathrm{T}} \sim 0.1$ the induced B-mode polarization ranges between $10^{-3}\, (\mu \mathrm{K})^2$ and 
$10^{-2}(\mu \mathrm{K})^2$ which is larger than what we have in Fig. \ref{Figure5} even assuming the 
maximum value of $\epsilon$. We must however bear in mind that the Faraday rate goes as $\nu^{-2}$ and, therefore, 
$C_{\ell}^{\mathrm{BB}}$ goes as $\nu^{-4}$. This means that by moving from $100$ GHz down to $30$ GHz the 
signal roughly increases by a factor $(10/3)^4\simeq 123$.

\renewcommand{\theequation}{5.\arabic{equation}}
\setcounter{equation}{0}
\section{Concluding remarks}
\label{sec5}
Often, in the literature, sharp and challenging statements on the interplay of inflationary magnetic fields and CMB anisotropies are not corroborated by detailed analyses. We tried to explore here the opposite perspective by following a more pedantic approach 
involving the initial conditions of the Einstein-Boltzmann hierarchy in the presence of inflationary seeds. According to some, 
the really important problem is to tailor specific magnetogenesis mechanisms with very little attention 
to potentially interesting observational consequences. Some others prefer to assume the existence of large-scale 
magnetic fields and to scrutinize their phenomenological signatures without bothering to ask why those fields exist.
As it has been shown in this paper, the two aforementioned perspectives must be seriously considered as complementary even if they 
are sometimes viewed as mutually exclusive. 

The necessity of bridging the gap between top-down and bottom-up approaches is, fortunately or unfortunately, inherent in all 
the analyses addressing the interplay of gravitational and gauge interactions in the early Universe. To be fair the same 
kind of problems emerge in the concordance lore. In this context concordance simply means a minimalistic 
agreement on the standard completion of the $\Lambda$CDM paradigm where the reheating is assumed to be almost sudden, 
inflation is driven by a single field and the post-inflationary history does not include long phases very different from radiation. 
Model independent analyses of the large-scale data are desirable but virtually 
impossible. In this spirit the initial conditions of the Einstein-Boltzmann hierarchy have been derived under 
the hypothesis that large-scale magnetic fields of inflationary origin were present prior to equality and for 
typical wavelengths larger than the Hubble radius at the corresponding epoch. 
Following earlier attempts, the guiding goal of the present investigation is to bring the treatment of magnetized CMB anisotropies to the same standards which are typical of the cases where large-scale magnetic fields are absent. 

The main assumption has been that the $\Lambda$CDM scenario with standard thermal history and inflationary completion is a sound approximation to the early dynamics of our Hubble patch. Here the amplitude of the magnetic fields 
is not given extrinsically as a further parameter but it depends on the standard $\Lambda$CDM 
parameters. The parameters of the magnetic power spectrum and hence 
of the CMB observables only depend on the slow-roll parameter.
As an application we presented the explicit computations of the temperature autocorrelations, 
of the polarization autocorrelations and of the cross-correlation power spectra of temperature and polarization. 
B-mode autocorrelations are potentially generated by Faraday rotation of the CMB whose 
linear polarization is  affected, in turn,  by the presence of the magnetic fields. 

Let us conclude with a conjecture. The $\Lambda$CDM scenario with tensor completion leads to an upper limit on 
the tensor to scalar ratio $r_{\mathrm{T}}$. Such a limit, by appropriately combining various data 
sets ranges from $r_{\mathrm{T}} <0.3$ down to $r_{\mathrm{T}}< 0.1$. The limit on $r_{\mathrm{T}}$ can be 
easily translated in a limit on the slow-roll parameter $\epsilon$ and then we discover for instance that if $r_{\mathrm{T}}< 0.12$ the 
modifications induced by the inflationary seed on the temperature and polarization power spectra are indistinguishable from the differences associated with the use of different data bases.  Conversely, if we believe that relic magnetic fields are there (and tensors are absent), then the present results suggest that  $\epsilon =0.03$ would already induce observable differences in the CMB spectra and this will imply an independent bound on $\epsilon$ possibly achievable with more accurate analyses.  In the near future less conventional models of inflationary magnetogenesis can be analyzed by using 
the same approach developed here.

\newpage
\begin{appendix}
\renewcommand{\theequation}{A.\arabic{equation}}
\setcounter{equation}{0}
\section{Correlation functions}
\label{APPA}
The comoving electric and magnetic fields are defined as 
\begin{equation}
\vec{E} = a^2 \, \sqrt{\lambda} \,\,\vec{e}, \qquad \vec{B} = a^2 \, \sqrt{\lambda} \,\, \vec{b}. 
\end{equation}
The fields $\vec{e}$ and $\vec{b}$ are introduced from the 
corresponding field strengths, i.e. $Y_{i\,0} = - a^2 \, e_{i}$ and $Y_{i\,j} = - a^2 
\epsilon_{i\, j\, k} \, b^{k}$.  The gauge action is canonical in terms of $\vec{E}$ and $\vec{B}$ and not in terms 
of $\vec{e}$ and $\vec{b}$. The evolution equations of the canonical modes derived 
from the action and their explicit form is:
\begin{eqnarray}
&& \frac{1}{\sqrt{\lambda}} \vec{\nabla} \cdot ( \sqrt{\lambda} \, \vec{E}) = 0,\qquad 
\sqrt{\lambda} \vec{\nabla} \cdot \biggl( \frac{\vec{B}}{\sqrt{\lambda}}\biggr) =0,
\label{EB1}\\
&& \frac{1}{\sqrt{\lambda}} \vec{\nabla} \times (\sqrt{\lambda} \, \vec{B} ) = \vec{J} + \frac{1}{\sqrt{\lambda}} \frac{\partial}{\partial \tau}( \sqrt{\lambda} \, \vec{E}), 
\label{EB2}\\
&& \sqrt{\lambda} \vec{\nabla} \times \biggl(\frac{\vec{E}}{\sqrt{\lambda}} \biggr) = 
- \sqrt{\lambda}  \frac{\partial}{\partial \tau} \biggl(\frac{\vec{B}}{\sqrt{\lambda}}\biggr),
\label{EB3}
\end{eqnarray}
where the possible presence of the Ohmic current has been included for completeness even if conducting initial conditions will not be considered explicitly. 
The system of Eqs. (\ref{EB1})--(\ref{EB3}), in the absence of electromagnetic sources, is invariant under the generalized duality transformation $\vec{E} \to - \vec{B}$, 
$\vec{B} \to \vec{E}$ and $\sqrt{\lambda} \to 1/\sqrt{\lambda}$ \cite{duality1,duality2}.
The conventions for the Fourier transform are:
\begin{equation}
B_{i}(\vec{x},\tau) = \frac{1}{(2\pi)^{3/2}} \int d^{3} k \, B_{i}(\vec{k},\tau) \, e^{- i \vec{k}\cdot\vec{x}}, \qquad 
E_{i}(\vec{x},\tau) = \frac{1}{(2\pi)^{3/2}} \int d^{3} k \, E_{i}(\vec{k},\tau) \, e^{- i \vec{k}\cdot\vec{x}}.
\label{FC}
\end{equation}
Consequently the fluctuations of the magnetic and electric energy densities is given by: 
\begin{eqnarray}
&& \delta \rho_{\mathrm{B}} (\vec{q},\tau) = \frac{1}{(2\pi)^{3/2}\,\, 8\pi a^4}\, \int d^{3} k\, \biggl[ B_{i}(\vec{k},\tau) B_{i}(\vec{q}- \vec{k},\tau) - \frac{4\pi^2}{k^3} \, P_{\mathrm{B}}(k,\tau) \delta^{(3)}(\vec{q})\biggr],
\nonumber\\
&& \delta \rho_{\mathrm{E}} (\vec{q},\tau) = \frac{1}{(2\pi)^{3/2}\,\, 8\pi a^4}\, \int d^{3} k\, \biggl[ E_{i}(\vec{k},\tau) E_{i}(\vec{q}- \vec{k},\tau) - \frac{4\pi^2}{k^3} \, P_{\mathrm{E}}(k,\tau) \delta^{(3)}(\vec{q})\biggr].
\label{en2}
\end{eqnarray}
The electric and magnetic anisotropic stresses are defined as
\begin{eqnarray}
\Pi^{(B)}_{ij}(\vec{q},\tau) &=&  \frac{1}{4\pi a^4 }\, \int \frac{d^{3} k}{(2\pi)^{3/2}}\, \biggl[ B_{i}(\vec{k},\tau) B_{j}(\vec{q}- \vec{k},\tau) - \frac{\delta_{ij}}{3} B_{m}(\vec{k},\tau) B_{m}(\vec{q} -\vec{k},\tau)\biggr],
\nonumber\\
\Pi^{(E)}_{ij}(\vec{q},\tau) &=&  \frac{1}{4\pi a^4 }\, \int \frac{d^{3} k}{(2\pi)^{3/2}} \, \biggl[ E_{i}(\vec{k},\tau) E_{j}(\vec{q}- \vec{k},\tau) - \frac{\delta_{ij}}{3} E_{m}(\vec{k},\tau) E_{m}(\vec{q} -\vec{k},\tau)\biggr].
\label{en5}
\end{eqnarray}
The stochastic averages of the fluctuations variables defined in Eqs. (\ref{en2})--(\ref{en5}) are all vanishing, i.e. using Eqs. (\ref{st1})--(\ref{st2}), $\langle\delta \rho_{\mathrm{B}}(\vec{x},\tau)\rangle=0$ and 
$\langle\delta \rho_{\mathrm{E}}(\vec{x},\tau)\rangle=0$ and similarly for the anisotropic stresses.
The second order correlations of the magnetic energy density and of the anisotropic stress are 
\begin{eqnarray}
&& \langle \delta\rho_{\mathrm{B}}(\vec{q},\tau) \,\delta\rho_{\mathrm{B}}(\vec{p},\tau) \rangle = \frac{2\pi^2}{q^3} {\mathcal Q}_{\mathrm{B}}(q,\tau) \, \delta^{(3)} (\vec{q} + \vec{p}),
\label{en7}\\
&& \langle \Pi_{\mathrm{B}}(\vec{q},\tau) \,\Pi_{\mathrm{B}}(\vec{p},\tau) \rangle = \frac{2\pi^2}{q^3} {\mathcal Q}_{\mathrm{B}\Pi}(q,\tau) \, \delta^{(3)} (\vec{q} + \vec{p}),
\label{en10}
\end{eqnarray}
where 
\begin{eqnarray}
&& {\mathcal Q}_{\mathrm{B}}(q,\tau) = \frac{q^{3}}{128\, \pi^3\, a^{8}}  \int 
d^{3} k \frac{P_{\mathrm{B}}(k,\tau)}{k^3} \frac{P_{\mathrm{B}}(|\vec{q} - \vec{k}|,\tau)}{|\vec{q} - \vec{k}|^3}\, \Lambda_{\rho}(k,q),
\label{en11}\\
&& {\mathcal Q}_{\mathrm{B}\Pi}(q,\tau) = \frac{q^3}{288\, \pi^3\, a^{8}(\tau) }
\int d^{3} k \frac{P_{\mathrm{B}}(k,\tau)}{k^3} \frac{P_{\mathrm{B}}(|\vec{q} - \vec{k}|,\tau)}{|\vec{q} - \vec{k}|^3}\,\Lambda_{\Pi}(k,q).
\label{en13}
\end{eqnarray}
The functions $\Lambda_{\rho}(k, q)$ and $\Lambda_{\Pi}(k,q)$ are defined as
\begin{eqnarray}
\Lambda_{\rho}(k,q) &=& 1 + \frac{[\vec{k} \cdot (\vec{q} - \vec{k})]^2}{k^2 |\vec{q} - \vec{k}|^2},
\label{en15}\\
\Lambda_{\Pi}(k,q) &=& 1 + \frac{[\vec{k}\cdot (\vec{q} - \vec{k})]^2}{k^2 |\vec{q} - \vec{k}|^2}  
+\frac{6}{q^2} \biggl[ \vec{k}\cdot( \vec{q} - \vec{k}) - \frac{[\vec{k}\cdot (\vec{q} - \vec{k})]^3}{k^2 | \vec{q} - \vec{k}|^2} \biggr]
\nonumber\\
&+& \frac{9}{q^4} \biggl[ k^2 |\vec{q} - \vec{k}|^2 - 2 [ \vec{k}\cdot(\vec{q} - \vec{k})|]^2 + 
\frac{[\vec{k}\cdot (\vec{q} - \vec{k})]^4}{k^2 |\vec{q} - \vec{k}|^2} \biggr].
\label{en16}
\end{eqnarray}
The functions $\Lambda_{\rho}(k,q)$ and $\Lambda_{\Pi}(k,q)$ coincide for magnetic and electric degrees of freedom 
since both $\vec{E}$ and $\vec{B}$ are solenoidal fields: $\vec{B}$ is solenoidal because of the absence of magnetic monopoles 
while $\vec{E}$ is solenoidal because the pprotoinflationary plasma is globally neutral and any electric charge asymmetry is absent. The explicit expressions of the power spectra of Eqs. (\ref{en11}) and (\ref{en13}) is  obtained by using the power spectra of Eqs. (\ref{PB})--(\ref{PE}).  

For typical wavelengths larger than the Hubble radius the second-order spectra including the slow roll corrections are given by:
\begin{equation}
{\mathcal Q}_{\mathrm{B}}(q,\tau) = {\mathcal O}_{\mathrm{B}}(q,\epsilon,f)\, \biggl(\frac{a}{a_{ex}}\biggr)^{g_{B}(\epsilon,f)},\quad 
{\mathcal Q}_{\mathrm{B}\Pi}(k,\tau) = {\mathcal O}_{\mathrm{B}\Pi}(q,\epsilon,f)\, \biggl(\frac{a}{a_{ex}}\biggr)^{g_{B}(\epsilon,f)},
\label{secord10}
\end{equation}
where $g_{B}(\epsilon,f) = 4 f - 8 + 4 \epsilon\,f$. The $k$-dependent amplitudes appearing in Eqs. (\ref{secord10}) 
are:
\begin{eqnarray}
&& {\mathcal O}_{\mathrm{B}}(q,\epsilon,f) = H^8 \,{\mathcal C}_{\mathrm{B}}(f,\epsilon)\, {\mathcal L}_{\mathrm{B}}(f, \epsilon, q) \biggl(\frac{q}{q_{\mathrm{p}}}\biggr)^{m_{\mathrm{B}}(\epsilon,f) -1}, 
\nonumber\\
&&{\mathcal O}_{\mathrm{B}}(q,\epsilon,f) = H^8 \,{\mathcal C}_{\mathrm{B}\Pi}(f,\epsilon)\, {\mathcal L}_{\mathrm{B}\Pi}(f, \epsilon, q) \biggl(\frac{q}{q_{\mathrm{p}}}\biggr)^{m_{\mathrm{B}\Pi}(\epsilon,f) -1},
\label{secord13}
\end{eqnarray}
where $m_{\mathrm{B}}(\epsilon,f)\, = m_{\mathrm{B}\Pi}(\epsilon,f) = 9 - 4 f( 1 + \epsilon)$. 
The functions ${\mathcal C}_{\mathrm{B}}(f,\epsilon)$  and ${\mathcal C}_{\mathrm{B}\Pi}(f,\epsilon)$ are given, respectively, by:
\begin{equation}
{\mathcal C}_{\mathrm{B}}(f,\epsilon) = \frac{2^{4 f( 1 + \epsilon)}}{1024\, \pi^7} \, \Gamma^4[f ( 1 + \epsilon) +1/2],\quad 
{\mathcal C}_{\mathrm{B}\Pi}(f,\epsilon) =\frac{4}{9} {\mathcal C}_{\mathrm{B}}(f,\epsilon).
\label{secord15}
\end{equation}
 The functions ${\mathcal L}_{\mathrm{B}}(f, \epsilon, q)$ and ${\mathcal L}_{\mathrm{B}\Pi}(f, \epsilon, q)$ are:
  \begin{eqnarray}
 {\mathcal L}_{\mathrm{B}}(f, \epsilon, q) &=& \frac{8[ f (1 + \epsilon) +1]}{3 [ 
 4f ( 1 + \epsilon) - 5][ 4 - 2 f ( 1 + \epsilon)]} - \frac{8}{3[ 4 - 2 f ( 1 + \epsilon)]} \biggl( \frac{q}{q_{0}}\biggr)^{ 2 f ( 1 + \epsilon)-4} 
\nonumber\\ 
&+& \frac{4}{5 - 4 f ( 1 + \epsilon)} \biggl( \frac{q}{q_{\mathrm{max}}}\biggr)^{ 4f ( 1 + \epsilon) -5},
  \label{secord17}\\
{\mathcal L}_{\mathrm{B}\Pi}(f, \epsilon, q) &=& \frac{2[ 17 -2 f (1 + \epsilon)]}{15 [ 
 4f ( 1 + \epsilon) - 5][ 4 - 2 f ( 1 + \epsilon)]} - \frac{2}{3[ 4 - 2 f ( 1 + \epsilon)]} \biggl( \frac{q}{q_{0}}\biggr)^{ 2 f ( 1 + \epsilon)-4} 
\nonumber\\
&+& \frac{7}{5 - 4 f ( 1 + \epsilon)} \biggl( \frac{q}{q_{\mathrm{max}}}\biggr)^{ 4f ( 1 + \epsilon) -5},
  \label{secord18}
 \end{eqnarray}
 The comoving scale $q_{\mathrm{p}} =0.002\, \mathrm{Mpc}^{-1}$ is the usual pivot scale at which the power spectra of the scalar curvature are assigned. The value of $q_{0}$ has been chosen $0.001\, q_{\mathrm{p}}$ while $q_{\mathrm{max}}$ 
 is related to the maximal amplified frequency of the magnetic field spectrum.
 \renewcommand{\theequation}{B.\arabic{equation}}
\setcounter{equation}{0}
\section{Evolution equations in the $\alpha$ parametrization.}
\label{APPB}
In the $\alpha$-parametrization the Hamiltonian and the momentum constraints read, respectively, 
\begin{eqnarray}
&& \frac{\partial h}{\partial\alpha} = \frac{\kappa^2 \alpha}{2 (\alpha+ 1)} \xi - \frac{3}{\alpha}
\biggl[\Omega_{\mathrm{R}}\biggl(R_{\nu} \delta_{\nu} + R_{\gamma} \delta_{\gamma}\biggr) + \Omega_{\mathrm{M}} 
\biggl( \frac{\Omega_{\mathrm{c}0}}{\Omega_{\mathrm{M}0}}
\delta_{\mathrm{c}} + \frac{\Omega_{\mathrm{b}0}}{\Omega_{\mathrm{M}0}}
\delta_{\mathrm{b}}\biggr)\biggr],
\label{eq6}\\
&& \kappa^2 \alpha^2 \frac{\partial \xi}{\partial \alpha} = - \frac{4}{\sqrt{\alpha+1}} \biggl\{ R_{\nu} \theta_{\nu} 
+ R_{\gamma} [ 1 + R_{\mathrm{b}}(\alpha)] \theta_{\gamma\mathrm{b}} + \frac{3}{4} \frac{\Omega_{\mathrm{c}0}}{\Omega_{\mathrm{M}0}} \alpha \theta_{\mathrm{c}}\biggr\}.
\label{eq7}
\end{eqnarray}
where $R_{\mathrm{b}}(\alpha)$ denote the baryon-to-photon ratio $R_{\mathrm{b}}(\alpha) 
= 3(\Omega_{\mathrm{b}0}/\Omega_{\mathrm{M}0}) \, \alpha/(4 R_{\gamma})  \simeq 0.215 \, \alpha$.
The remaining two equations stemming from the perturbed Einstein equations can be written as:
\begin{eqnarray}
&& \frac{\partial^2 h}{\partial \alpha^2} + \frac{5 \alpha + 4}{2 \alpha (\alpha +1)} \frac{\partial h}{\partial\alpha} - \frac{\kappa^2 \xi}{2 (\alpha + 1)}= \frac{3}{\alpha^2 (\alpha+1)}\biggl[ R_{\gamma} \delta_{\gamma} + R_{\nu} \delta_{\nu} + R_{\gamma} \Omega_{\mathrm{B}}\biggr],
\label{eq8}\\
&& \frac{\partial^2 {\mathcal Q}}{\partial \alpha^2} + \frac{5 \alpha + 4 }{2 \alpha(\alpha+1)} \frac{\partial {\mathcal Q}}{\partial \alpha}= 
\frac{\kappa^2 \xi}{2 (\alpha + 1)} + \frac{12}{\alpha^2 (\alpha +1)} (R_{\nu} \sigma_{\nu} 
+ R_{\gamma} \sigma_{\mathrm{B}}).
\label{eq9}
\end{eqnarray}
where ${\mathcal Q}= (h + 6 \xi)$; the evolution equations of the neutrinos obey, 
\begin{eqnarray}
&& \frac{\partial \delta_{\nu}}{\partial \alpha} = - \frac{2 \theta_{\nu}}{3 \sqrt{\alpha +1}} + \frac{2}{3} \frac{\partial h}{\partial \alpha}, 
\qquad \frac{\partial \theta_{\nu}}{\partial \alpha} = \frac{\kappa^2}{8 \sqrt{\alpha +1}} \delta_{\nu} 
- \frac{\kappa^2 }{2 \sqrt{\alpha +1}} \sigma_{\nu},
\label{eq10}\\
&& \frac{\partial {\sigma}_{\nu}}{\partial \alpha} = \frac{2 \theta_{\nu}}{15 \sqrt{\alpha + 1}} - \frac{2}{15} 
\frac{\partial {\mathcal Q}}{\partial \alpha}- \frac{3}{20} \frac{\kappa {\mathcal F}_{\nu\,3}}{\sqrt{\alpha +1}} ,
\label{eq10a}\\
&& \frac{\partial {\mathcal F}_{\nu\,\ell}}{\partial \alpha} = \frac{\kappa}{2 ( 2 \ell + 1)\sqrt{\alpha + 1}} 
[\ell {\mathcal F}_{\nu\, (\ell -1)} - (\ell +1) {\mathcal F}_{\nu\, (\ell +1)}],\qquad \ell \geq 3
\label{eq11}
\end{eqnarray}
The evolution equations of the dark-matter sector obey instead 
\begin{equation}
\frac{\partial \delta_{\mathrm{c}}}{\partial \alpha}= - \frac{\theta_{\mathrm{c}}}{2 \sqrt{\alpha +1}} + \frac{1}{2} \frac{\partial h}{\partial \alpha}, \qquad \frac{\partial \theta_{\mathrm{c}}}{\partial \alpha} + \frac{\theta_{\mathrm{c}}}{\alpha} =0
\label{eq12}
\end{equation}
In in the tight-coupling  approximation the equations of the baryon-photon system are:
\begin{eqnarray}
&&\frac{\partial \theta_{\gamma\mathrm{b}}}{\partial \alpha} + \frac{R_{\mathrm{b}}\,\theta_{\gamma\mathrm{b}}}{\alpha ( R_{\mathrm{b}} +1)}  = \frac{\kappa^2\,\delta_{\gamma}}{8 \sqrt{\alpha +1} (R_{\mathrm{b}} +1)} + \frac{\kappa^2\,(\Omega_{\mathrm{B}} - 4 \sigma_{\mathrm{B}})}{8 \sqrt{\alpha +1} (R_{\mathrm{b}} +1)},
\label{eq13}\\
&& \frac{\partial \delta_{\gamma}}{\partial \alpha} = - \frac{2}{3} \frac{\theta_{\gamma\mathrm{b}}}{\sqrt{\alpha +1}} + \frac{2}{3} \frac{\partial h}{\partial \alpha},\qquad 
\frac{\partial \delta_{\mathrm{b}}}{\partial \alpha} = - \frac{\theta_{\gamma\mathrm{b}}}{2 \sqrt{\alpha +1}} + \frac{1}{2} \frac{\partial h}{\partial \alpha}.
\label{eq14}
\end{eqnarray}
\end{appendix}

\end{document}